\documentclass[journal]{IEEEtran}

\usepackage{cite}
\usepackage{amsmath,amssymb,amsfonts}
\usepackage{algorithmic}
\usepackage{graphicx}
\usepackage{textcomp}

\usepackage{amsmath,graphicx}
\usepackage{amsfonts}
\usepackage{xcolor}
\usepackage{mathrsfs}
\usepackage{cite}
\usepackage{epstopdf}
\usepackage{color}
\usepackage{url}
\usepackage{booktabs}
\usepackage{multirow}

\usepackage{colortbl}
\definecolor{rgb6}{RGB}{255, 186, 119}
\definecolor{rgb3}{RGB}{158, 216, 229}
\definecolor{rgb9}{RGB}{160, 188, 33}

\def\etal{\emph{et al. }}
\def\eg{\emph{e.g. }}
\def\ie{\emph{i.e. }}

\begin{document}
\title{COVID-19 Chest CT Image Segmentation --\\ A Deep Convolutional Neural Network Solution}
\author{Qingsen Yan,
		Bo Wang,
		Dong Gong,
		Chuan Luo,
		Wei Zhao, \\
		Jianhu Shen,
		Qinfeng Shi,
        Shuo Jin,
		Liang Zhang
		and Zheng You
\thanks{The work is supported by Application for Independent Research Project of Tsinghua University (Project Against SARI), Zhejiang University special scientific research fund for COVID-19 preverntion and control, ARC (DP160100703).
Q. Yan, B. Wang, W. Zhao and J. Shen are with the Beijing Jingzhen Medical Technology Ltd. (e-mail: qingsenyan@gmail.com, wang-b17@mails.tsinghua.edu.cn, zhaowei729406@gmail.com, shenjainhu@jingzhentech.com).
B. Wang and Z. You are with the State Key Laboratory of Precision Measurement Technology and Instruments, Department of Precision Instrument; Innovation Center for Future Chips, Tsinghua University (THU). (e-mail: yz-dpi@mail.tsinghua.edu.cn).
S. Jin is with the Beijing Tsinghua Changgung Hospital, School of Clinical Medicine, THU. (e-mail:
jsa01263@btch.edu.cn)
D. Gong and Q. Shi are with the Australian Institute for Machine Learning, The University of Adelaide. (e-mail:
edgong01@gmail.com, shiqinfeng@gmail.com)
C. Luo is with State Key Laboratory of Precision Measurement Technology and Instruments, THU.(e-mail:
luochuan@mail.tsinghua.edu.cn)
L. Zhang is with the School of Computer Science and Technology, Xidian University. (e-mail: liangzhang@xidian.edu.cn)} 
\thanks{The first two authors are contributed equally to this work. Corresponding authors: S. Jin, L. Zhang and Z. You.}
}

\maketitle

\begin{abstract}
Objective: A novel coronavirus disease 2019 (COVID-19) was detected and has spread rapidly across various countries around the world since the end of the year 2019, Computed Tomography (CT) images have been used as a crucial alternative to the time-consuming RT-PCR test. However, pure manual segmentation of CT images faces a serious challenge with the increase of suspected cases, resulting in urgent requirements for accurate and automatic segmentation of COVID-19 infections. Unfortunately, since the imaging characteristics of the COVID-19 infection are diverse and similar to the backgrounds, existing medical image segmentation methods cannot achieve satisfactory performance. Methods: In this work, we try to establish a new deep convolutional neural network tailored for segmenting the chest CT images with COVID-19 infections. We firstly maintain a large and new chest CT image dataset consisting of 21,658 annotated chest CT images from 861 patients with confirmed COVID-19. Inspired by the observation that the boundary of the infected lung can be enhanced by adjusting the global intensity, in the proposed deep CNN, we introduce a feature variation block which adaptively adjusts the global properties of the features for segmenting COVID-19 infection. The proposed FV block can enhance the capability of feature representation effectively and adaptively for diverse cases. We fuse features at different scales by proposing Progressive Atrous Spatial Pyramid Pooling to handle the sophisticated infection areas with diverse appearance and shapes. Results: The proposed method achieves the state-of-the-art performance. Dice similarity coefficients are 0.987 and 0.726 for lung and COVID-19 segmentation, respectively. Conclusion: We conducted experiments on the data collected in China and Germany and show that the proposed deep CNN can produce impressive performance effectively. Significance: The proposed network enhances the segmentation ability of the COVID-19 infection, makes the connection with other techniques and contributes to the development of remedying COVID-19 infection.
\end{abstract}


\section{Introduction}
\label{sec-intro}
\IEEEPARstart{I}{n} December 2019, coronavirus disease 2019 (COVID-19) a new febrile respiratory tract illness caused by severe acute respiratory syndrome coronavirus 2 (SARS-CoV-2) was detected.
The typical onset symptoms of COVID-19 patients are fever, cough, myalgia, dyspnea, and muscle aches.
Despite the imposition of strict quarantine rule to limit its propagation, the COVID-19 infection has spread rapidly affecting countries worldwide.
At the end of January 2020, the World Health Organization (WHO) declared that COVID-19 becomes a Public Health Emergency of International Concern \cite{who}. 
As of 11 April 2020, the WHO reported 1,610,909 worldwide cases with 99,690 deaths \cite{whodata}.
While infection rates are decreasing in China, numbers of new infections are still exponentially growing in many other countries.

Reverse transcription polymerase chain reaction (RT-PCR) is one of the standard diagnostic methods to detect nucleotides from specimens obtained by oropharyngeal swab, nasopharyngeal swab, bronchoalveolar lavage, or tracheal aspirate \cite{Bai20Per}.
However, recent reports have indicated that the sensitivity of RT-PCR might not be high enough for detecting COVID-19 \cite{Tao20Cor,Fang20Sen}, which can possibly be attributed to quality, stability and insufficient viral material in specimens.
On the other hand, since chest Computed tomography (CT) images captured from COVID-19 patients frequently show bilateral patchy shadows or ground glass opacity in the lung \cite{Wang20clin}, CT has become a vital complementary tool for detecting the lung associated with COVID-19.
Comparing to RT-PCR test, chest CT is relatively easy to operate and has a high sensitivity for screening COVID-19 infection \cite{Tao20Cor}.
Therefore, CT could serve as a practical approach for early screening and diagnosis of COVID-19 in China.
However, as the increment of confirmed and suspected cases of COVID-19, manually contouring of lung lesions is a tedious and labor-intensive task.
To speed up diagnosis and improve access to treatment, developing a fast automatically segmentation for COVID-19 infection is critical for the disease assessment.

Recently, with the rapid development of artificial intelligence, \cite{yan2020deep,yan2019two,YANwacv2019,yan2019attention,Gong2016From,he19Knowledge,dong19Memorizing},
deep learning technology has been widely used in medical image processing due to its powerful feature representation.
Several techniques based on deep learning have published to detect COVID‐19 pneumonia from CT images \cite{wang2019covid, Wang2020A, Ayrton20using, Muhammad20Can}.
Wang \etal \cite{Wang2020A} developed a deep learning method that could extract COVID-19's graphical features in order to provide a clinical diagnosis ahead of the pathogenic test.
Ayrton \cite{Ayrton20using} adopted the transfer learning technique with ResNet50 backbone to detect COVID-19.
Wang \etal \cite{wang2019covid} introduced a deep convolutional neural network design tailored, called COVID-Net for the detection of COVID-19 cases from chest radiography images.
Gozes \etal \cite{Gozes2020rapid} presented a system that utilizes 2D and 3D deep learning models, modified and adapted existing deep network models and combined them with clinical understanding.
Tang \etal \cite{Tang2020Severity} trained a random forest (RF) model to assess the severity (non-severe or severe) based on quantitative features.
Shi \etal \cite{Shi2020large} proposed an infection Size Aware Random Forest method (iSARF) for classification.
Shan \etal \cite{Shan2020lung} developed a deep learning-based system for segmentation and quantification of infection regions from CT scans.
In summary, some deep learning based methods have been proposed to detect COVID-19 and viral pneumonia in chest CT images.
To our knowledge, however, only few publications have investigated the segmentation task for COVID-19 chest CT images.

\def \wid{8.6cm}
\begin{figure}[t]
\centering
\includegraphics[width=\wid]{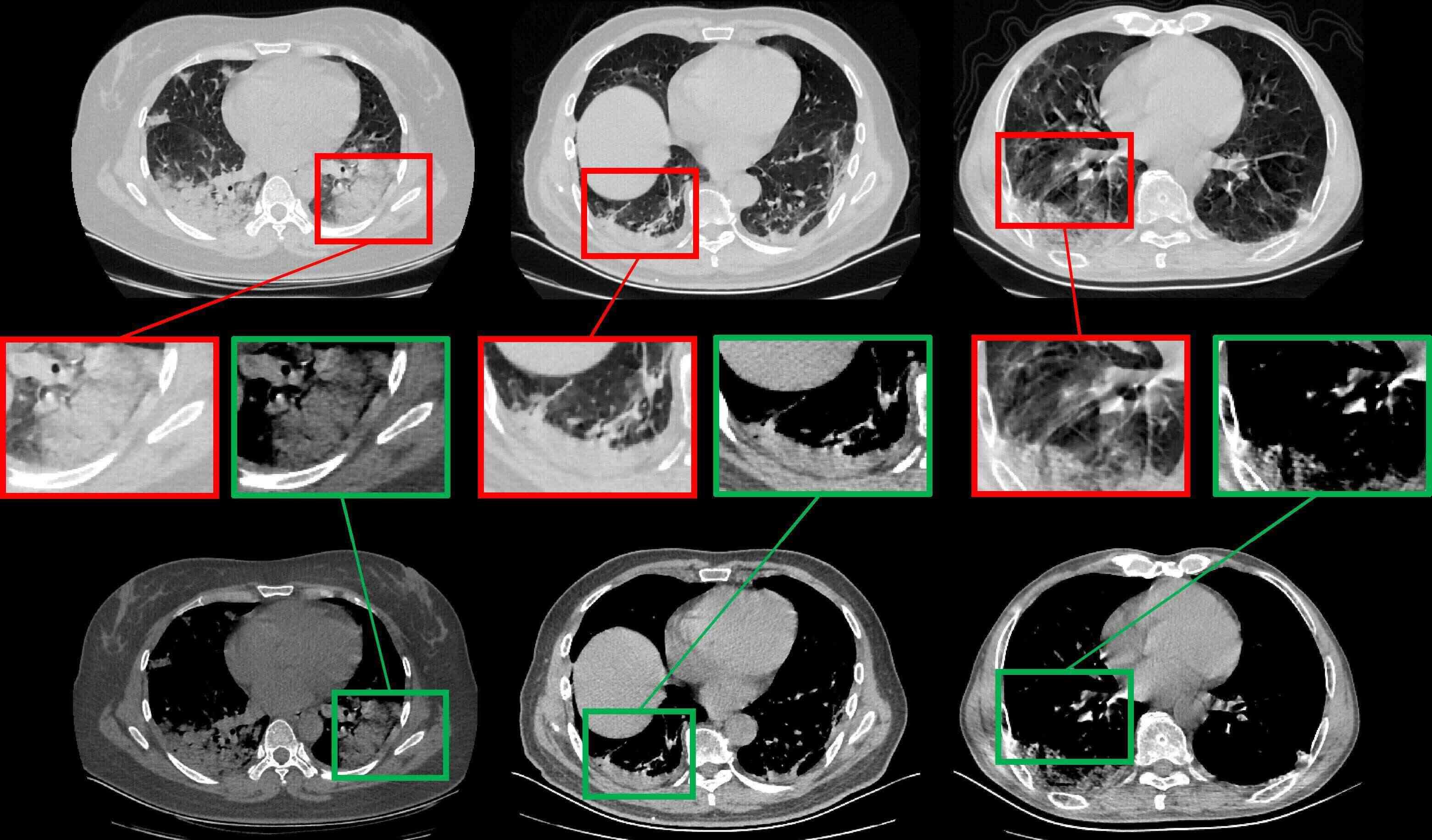}
\caption{Chest CT images of three patients with laboratory proven COVID-19 pneumonia. 
As shown in the top row, patchy ground-glass opacities (GGOs) and areas of consolidation bilaterally exist in all lung lobes (highlighted with red bounding box). It is hard to distinguish COVID-19 infection regions from the chest wall. 
The boundaries of COVID-19 infection regions are highlighted (as indicated by green bounding box), after carefully adjusting the window breadth and window locations for each CT image. 
}
\label{Fig_intro}
\end{figure}

In this paper, we try to establish a new tailored deep convolutional neural network (CNN) for segmenting the chest CT images with COVID-19 infections. 
Fig. \ref{Fig_intro} shows the chest CT images with COVID-19 infection, which contain ground-glass opacities (GGOs), areas of consolidation, and a mix of both in all lung lobes. 
Most lesions were located peripherally, with a slight preponderance of dorsal lung areas.
Due to the special structure and visual characteristics, the boundaries of COVID-19 infection regions are difficult to distinguish from the chest wall, making accurate segmentation for COVID-19 infection regions difficult. 
We observe that the boundaries of COVID-19 infection regions will be revealed by adjusting different parameters of window breadth and window locations in annotation processing, as shown in Fig. \ref{Fig_intro}, which can be beneficial for the COVID-19 infection image segmentation.

We propose a three-dimensional (3D) convolution based deep learning method for automatic segmentation of COVID-19 infection regions as well as the entire lung from chest CT images, referred to as COVID-SegNet. 
The proposed method can be hugely beneficial for the early screening of patients with COVID-19.
Inspired by the observation in annotation processing, the boundaries of COVID-19 infection regions are highlighted by adjusting the window breadth and window locations, we deign a Feature Variation (FV) block to handle the confusing boundaries.
The central idea of the FV block is to implicitly enhance the contrast and adjust the intensity in the feature level automatically and adaptively for different images. 
Based on the captured features of previous layers, the FV block employs channel attention to obtain the global parameter to generate new features.
In addition to the channel attention, the FV block uses spatial attention to guide the feature extraction from inputs in the encoder.
Aggregating these features can effectively enhance the capability of feature representation for the segmentation of COVID-19.
Furthermore, we propose a Progressive Atrous Spatial Pyramid Pooling (PASPP) to handle the challenging shape variations of COVID-19 infection areas.
PASPP consists of a base convolution module followed by a cascade of atrous convolutional layers, which uses multistage parallel fusion branches to obtain the final features.
Each atrous convolutional layer in PASPP only uses atrous filters with a reasonable dilation rate to cover different receptive fields.
And by the progressively aggregated information from atrous convolutional layers, the information from multiple scales is effectively fused, which further promotes the performance of COVID-19 pneumonia segmentation.

The main contributions of the paper can be summarized as:
\begin{itemize}
\item We propose a novel deep neural network (COVID-SegNet) for the segmentation of COVID-19 infection regions as well as the entire lung from chest CT images.

\item To address the key issue in the delineation of COVID-19 infection regions, a specific block, called Feature Variation (FV) block, is proposed to solve the problem of difficulty distinguishing COVID-19 pneumonia from the lung.

\item We introduce Progressive Atrous Spatial Pyramid Pooling (PASPP), which progressively aggregates information and obtains more effective contextual features.

\item To train the proposed networks, we maintain a novel and large dataset that consists of 21,658 chest CT images from 861 patients with confirmed COVID-19, which are annotated by experts. Ten cases captured from Germany are also used to test the robustness of the model.
\end{itemize}

\section{Materials}
\label{Sec-Mater}

\subsection{Dataset Introduction}
This study was approved by the medical ethics committees of the participating hospitals. Further consent was waived with approval.
In total, chest CT images of 861 patients with confirmed COVID-19 by RT-PCR are included in this study.
These CT images were acquired at 5 Chinese hospitals (Beijing Tsinghua Changgung Hospital, Wuhan No.7 Hospital, Zhongnan Hospital of Wuhan University, Tianyou Hospital Affiliated to Wuhan University of Science \& Technology, Wuhan's Leishenshan Hospital) between January 2 and February 26, 2020.
All imaging data were reconstructed by using a medium sharp reconstruction algorithm with a thickness of 0.625-10 mm (81\% under 2mm). 
To protect privacy, we deleted the personally identifiable information (PII) from all CT scans.
A total of 731 patient's CT images were randomly extracted for training. The remaining CT images of 130 patients were used as the testing set.

\subsection{Dataset Annotation}
Although we captured enough data of the COVID-19 chest CT images, accurate annotated labels are also indispensable. 
To enable the model to learn on accurate annotations, we build a team of six annotators with deep radiology background and proficient annotating skills to annotate the areas and boundaries of the lung and COVID-19 infection regions. 
Also, the quality of the final annotations is assessed by a senior radiologist with frontline clinical experience of COVID-19.

\def \wid{14.5cm}
\begin{figure*}[t]
\centering
\includegraphics[width=\wid]{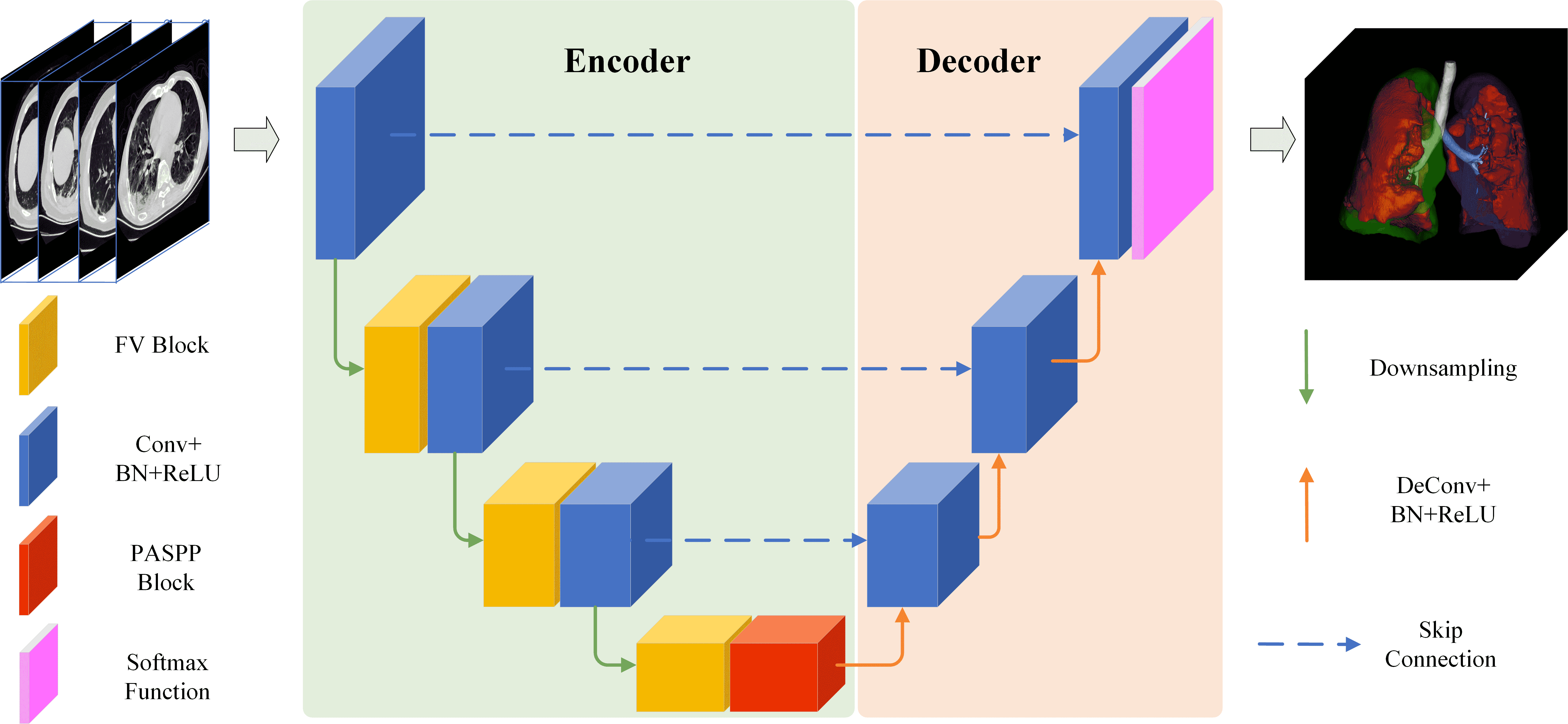}
\caption{The architecture of the proposed COVID-SegNet. The network includes encoder part for feature extraction and decoder part for estimating the segmentation results. The FV block is adopted to highlight contrast and position of COVID-19, the PASSP block is built based on progressively fusing the output of different arous convolutional layers.
The visualized final result is a presentation of the 3D segmentation of lung and the regions associated with COVID-19 infection.}
\label{Fig_frame}
\end{figure*}

\section{Method}
\label{sec-method}
In this section,  we start with the overview of the proposed approach, then introduce the feature variation block and progressive atrous spatial pyramid pooling block.
We briefly discuss the training strategy and implementation details in the end.

\subsection{Network Structure of COVID-SegNet}
We present a unified high-accuracy network for the segmentation of COVID-19 infection from chest CT images.
This network consists of two parts: Encoder and Decoder.
As shown in Fig. \ref{Fig_frame}, the encoder with 4 layers (\ie E1, E2, E3, E4) obtains robust information via feature extractor and PASPP.
Each layer employs residual and FV blocks as the basic operations for feature extractors, except the E4 layer.
The residual block adds up the input features and the results after two convolutional layers,  which effectively alleviates the vanishing gradient.
To preserve multiple contextual information and enlarge the receptive field, we use PASPP with different dilate rates on the final E4 layer.
After obtaining the encoded features, the decoder tries to restore the features to its original input size, which can remove the information loss induced by down-sampling from Encoder.
The decoder has three layers (D3, D2, D1). Each decoder layer allows the networks to gradually propagate the global contextual information to a higher resolution layer.
After a sigmoid activation function, we obtain the final segmentation of COVID-19 infection regions.
In addition, the skip connection is adopted to concatenate the output features of the encoder and input features of the decoder.
In this paper, the main contribution is we improve the encoder by adding \textit{FV block} and \textit{PASPP block} to better capture effective features. The overview of these two blocks is as follows.

We introduce the architectures of {FV block} by considering a material fact, the boundaries of COVID-19 infection regions are highlighted by adjusting the window breadth and window locations.
As shown in Fig. \ref{Fig_fv}, the proposed FV block includes three branches, \eg contrast enhancement branch, position sensitive branch, identity branch, which can automatically change the parameter to display the boundaries and position of COVID-19.
Specifically, the contrast enhancement branch learns a global parameter via a channel attention unit to highlight useful boundary information.
The position sensitive branch obtains a weight map by spatial attention unit to focus on the COVID-19 regions.
Finally, the FV block preserves more useful information by fusing these refined features.

The PASPP block takes the featured extracted with FV block as input and acquires semantic information with different receptive fields showing in Fig. \ref{Fig_paspp}.
Although ASPP has been proposed to capture global information for semantic segmentation, we claim that aggregating information progressively is a more reasonable approach to get effective features.
The PASPP block adopts atrous convolutions with different dilation rates to obtain features with various scales.
The final output is generated straightforwardly to assemble residual branches in parallel.

\def \wid{15cm}
\begin{figure*}[t]
\centering
\includegraphics[width=\wid]{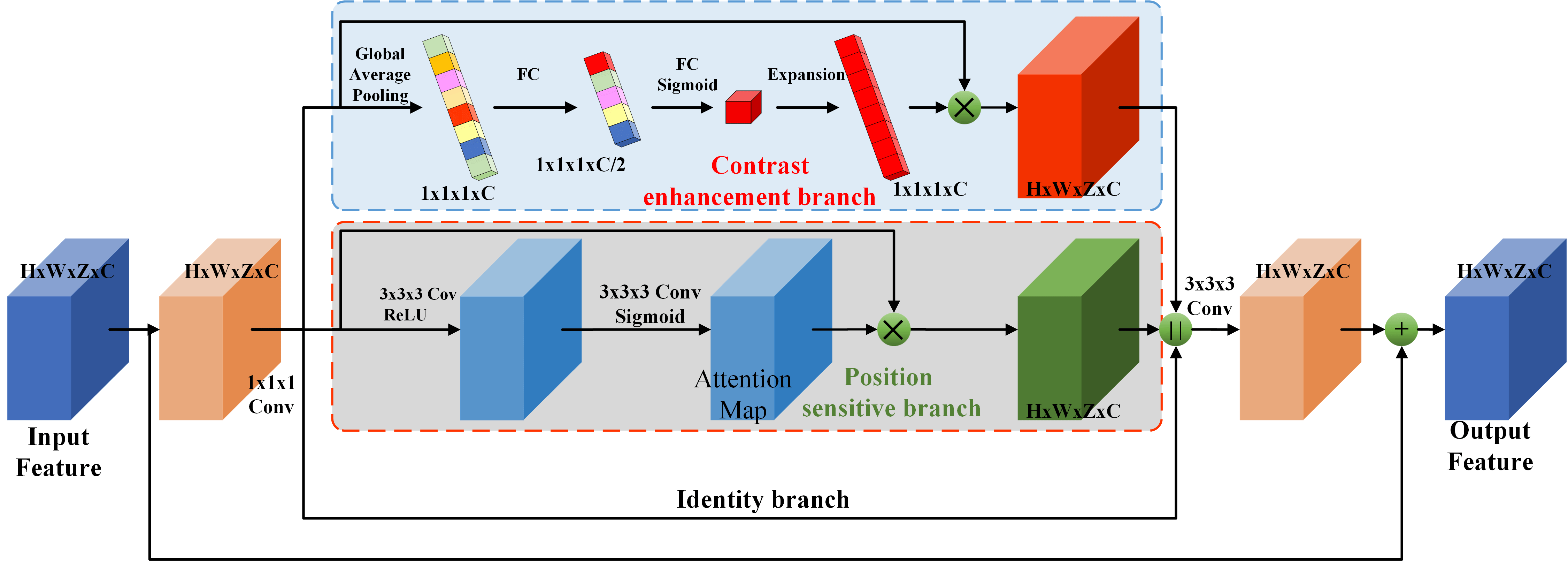}
\caption{FV block consists of contrast enhancement branch, position sensitive branch, and identity branch. The features of these branches are concatenated to decrease the number of the channel via a $3 \times 3 \times 3$ convolutional layer. The output features are obtained after residual learning with input.
}
\label{Fig_fv}
\end{figure*}

\subsection{Feature Variation}
As mentioned before, the boundaries of COVID-19 infection regions are highlighted by adjusting the window breadth and window locations.
In Fig. \ref{Fig_fv}, the designed FV block, which includes contrast enhancement branch, position sensitive branch and identity branch, tries to enhance the contrast of features and highlight the useful regions.
Let $Fv_{in}$ denotes the input feature, the features after $1 \times 1 \times 1$ represent $Fv_1$.
The output feature $Fv_{out}$ is given as,
\begin{equation}
\label{eq-fv}
Fv_{out} = Fv_{in} + Cov_{3}(Conca(C(Fv_1), P(Fv_1), Fv_1)),
\end{equation}
where $Cov_{3}(\cdot)$ denotes the $3 \times 3 \times3 $ convolutional layer, $Conca(\cdot)$ is the concatenation operation, $C(\cdot)$ represents the contrast enhancement branch, $P(\cdot)$ is the position sensitive branch.
The form of residual learning in Eq. (\ref{eq-fv}) implies that the information from the early blocks can quickly flow to the later blocks, and the gradient can be quickly back-propagated to the early blocks from the later blocks \cite{zhao16deep}.
The details of each sub-module are as follows.

\subsubsection{Contrast Enhancement Branch}
To enhance the contrast of features, the contrast enhancement branch $Con(\cdot)$ in Eq. (\ref{eq-fv}) attempts to learn a global parameter $F_g$ for input feature $Fv_1$ (See Fig. \ref{Fig_fv}).
The corresponding function is given as,
\begin{equation}
F_g = Cov_{1}(Cov_{3}(GAP(Fv_1))),
\end{equation}
where $Cov_{1}(\cdot)$ denotes the $1 \times 1 \times1 $ convolutional layer, $GAP(\cdot)$ represents global average pooling.
The values of $F_g$ is in the range $[0,1]$.
We obtain a channel weight map $F'_g$ via expansion, thus the number of $F'_g$ is consistent with $Fv_1$.
Finally, the output of contrast enhancement branch $F_c$ can be formulated as below,
\begin{equation}
F_C = F'_g \otimes Fv_1，
\end{equation}
where $\otimes$ denotes the element-wise multiplication.
Note that, instead of calculating a sequence of weight for feature $Fv_1$, we generate one weight for all the features of $Fv_1$.
This process is exactly corresponding to adjust the window breadth and window locations. Thus we deem it has the ability to generate enhanced features.

\subsubsection{Position Sensitive Branch}
The goal of position sensitive branch is to discard harmful information and highlight the helpful features, which are used to segmentation of COVID-19 infection. 
This branch $P(\cdot)$ in Eq. (\ref{eq-fv}) is a small network.
The architecture of position sensitive branch is displayed in Fig. \ref{Fig_fv}.
The attention map $A$ is calculated using input feature $Fv_1$ after two convolutional layers. 
Each layer adopts $3 \times 3\times 3$ convolution. The two convolutional layers are followed by a ReLU function and a sigmoid function, respectively. In the end, the output of this branch $F_P$ is obtained by element-wise multiplication between $Fv_1$ and the attention map.
\begin{equation}
F_P = A \otimes Fv_1.
\end{equation}
The values in $A$ are still in the range $[0, 1]$.
The attention map has same size as input feature.

\subsection{Progressive Atrous Spatial Pyramid Pooling}
In this subsection, we start with preliminary knowledge of atrous spatial pyramid pooling, then introduce the proposed PASPP block.

\subsubsection{Atrous Spatial Pyramid Pooling}
Global information captured by a large receptive field is essential for medical semantic segmentation.
To increase the receptive field size and decrease the number of convolutional layers, arous convolution is first proposed in \cite{chen16deeplab} to obtain enough global information while keeping the size of the feature map unchanged.
In one dimensional case, let $y[i]$ represents output and $x[i]$ denotes input, atrous convolution can be formulated as follows:
\begin{equation}
y[i] = \sum_{k=1}^{K}x{[i+d \cdot k]} \cdot w[k],
\end{equation}
where $K$ denotes the filter size, $d$ represents the dilation rate, and $w[k]$ is the $k$-th parameter of filter.
A larger dilation rate will capture a larger receptive field.
To produce different receptive fields, atrous spatial pyramid pooling taking atrous convolutions with different dilation rates to generate features with various scales. 
These features are concatenated together. Thus the outputs are indeed a sampling of the input with different scales information. 

\def \wid{9cm}
\begin{figure}[t]
\centering
\includegraphics[width=\wid]{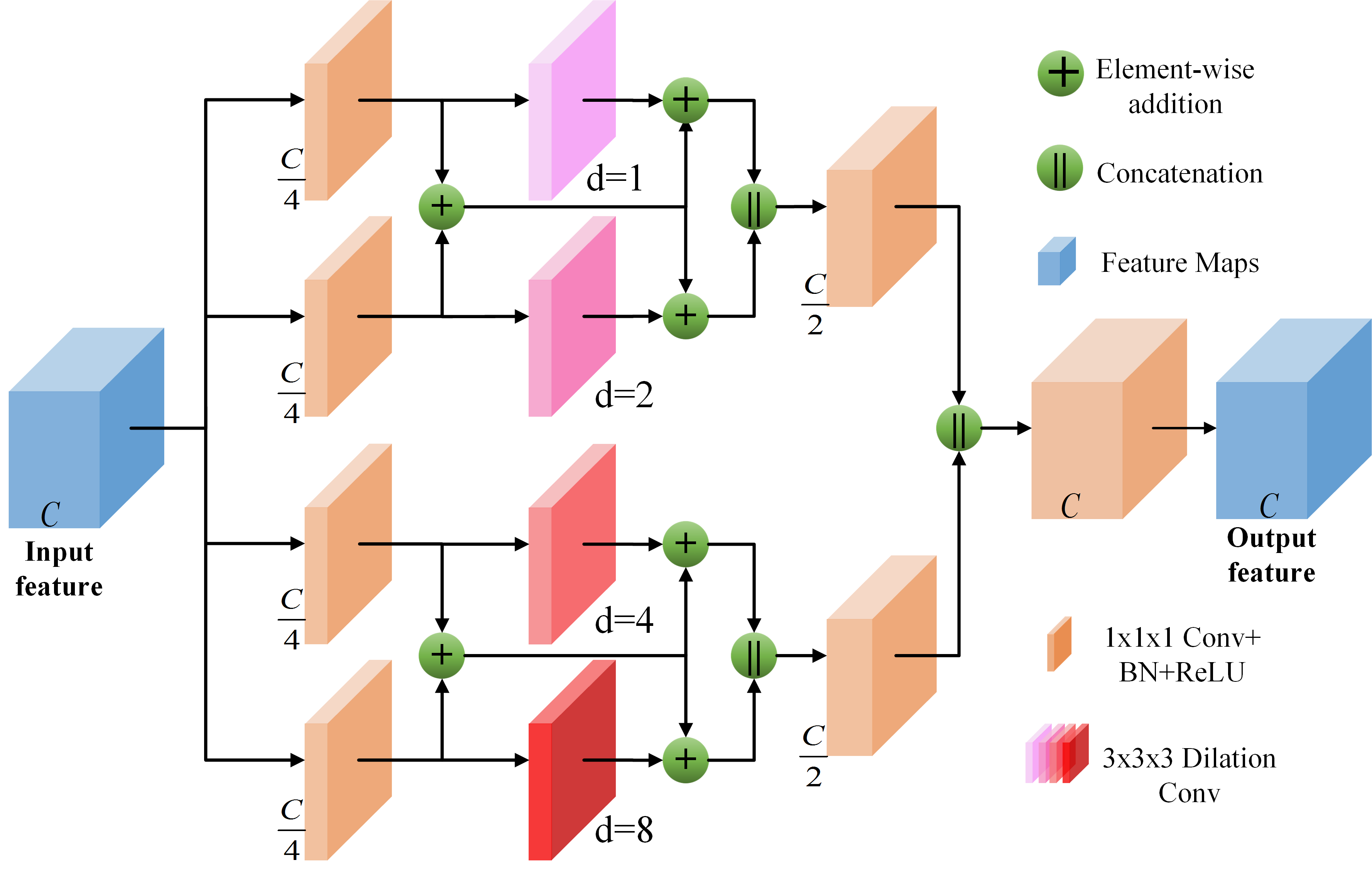}
\caption{The structure of PASPP block. We assemble two residual branches in parallel and sum up the outputs from two $1 \times 1 \times 1$ convolutional layers, then the outputs of two branches are progressively blended. Note that, compared to input features, the number of the channel decreases to quarter after each $1 \times 1 \times 1$ convolutional layer.
}
\label{Fig_paspp}
\end{figure}

\subsubsection{The PASPP Block}
In COVID-19 segmentation task, the infection regions often have very different sizes (See Fig. \ref{Fig_intro}).
To alleviate this dilemma, the features must be able to include different receptive fields.
For this goal, we employ ASPP in our network and progressively fuse the features with different receptive fields.
The structure of PASPP is illustrated in Fig. \ref{Fig_paspp}.
Given the input feature of PASPP $Fp_{in}$, we obtain four features $Fp_1, Fp_2, Fp_3, Fp_4$ by four $1 \times 1 \times 1$ convolutional layers in parallel.
Note that, compared to input features, the number of the channel decreases to quarter after each $1 \times 1 \times 1$ convolutional layer (See the second column in Fig. \ref{Fig_paspp}).
Then each branch feeds the feature into different atrous convolutional layer, respectively. The corresponding function is given as,
\begin{eqnarray}
Fd_t = Cov_{3}^{d}(Fp_t), \quad t= 1,2,3,4; d=2^{t-1},
\end{eqnarray}
where $Cov_{3}^{d}$ denotes the $3 \times 3 \times 3$ atrous convolutional layer with dilation rate $d$, $Fd_t$ represents the output feature of the $i$-th branch after $Cov_{3}^{d}$.
Sum the inputs of two adjacent atrous convolution branches, and add the sum to the output of each residual branch as the input of the subsequent layer. It is formulated as below,
\begin{eqnarray}
\begin{cases}
Fd'_t = Fd_t + Fd_1 + Fd_2, \quad t= 1,2\\
Fd'_t = Fd_t + Fd_3 + Fd_4 , \quad t= 3,4.
\end{cases}
\end{eqnarray}
where $Fd'_t$ denotes the output features of $t$-th branch.
To get effective features, $Fd'_t, t=1,2,3,4$ will be progressively aggregated based on adjacent features in parallel.
\begin{eqnarray}
\begin{cases}
Fd''_{1} = Cov_{1}(Conca(Fd'_1, Fd'_2))\\
Fd''_{2} = Cov_{1}(Conca(Fd'_3, Fd'_4))
\end{cases}
\end{eqnarray}
The $Fd''_{1}$ tends to fuse the information with small receptive field, $Fd''_{2}$ prones to capture features with larger receptive field. The channel's number of $Fd''_{1}$ and $Fd''_{2}$ is half of the input feature.
All the information are assembled by:
\begin{eqnarray}
Fp_{out} = Cov_{1}(Conca(Fd''_1, Fd''_2)),
\end{eqnarray}
where $Fp_{out}$ denotes the output features of PASPP block.

\section{Experiments}

\subsection{Dataset}
The dataset used in this study consists of 21,658 annotated chest CT images, with 861 patients confirmed COVID-19.
A total of 731 patient's CT images are randomly extracted for training. The remaining CT images of 130 patients are used as the testing set.

\subsection{Evaluation Metrics}
The screening performance of the proposed method is conducted by the Dice similarity coefficient, sensitivity, and precision.
The Dice similarity coefficient (Dice) represents a similarity metric between the ground truth, and the prediction score maps \cite{milletari2016v}.
It is calculated as follows:
\begin{eqnarray}
\label{eq-dic}
Dic(A, B) = \frac{2|A\cap B|}{|A|+|B|},
\end{eqnarray}
where $A$ is the segmented infection region, $B$ denotes the corresponding reference region, $|A\cap B|$ represents the number of pixels common to both images.
Sensitivity denotes the number of correctly identified positives with respect to the number of positives.
Precision is the fraction of positive instances among the retrieved instances.

\subsection{Implementation Details}

\textbf{The parameters of the network.} For the proposed framework, the encoding layers are residual blocks, FV blocks, PASSP blocks, and downsampling, while the decoding layers are residual blocks and deconvolution layers kernels with a stride of 1/2. The last layer is a softmax activation function to produce the segmentation results. 
All layers use $3 \times 3 \times 3$ kernels, if not specified otherwise.
Each convolutional layer is followed by batch normalization and ReLU.
The channel numbers are doubled each layer from 64 to 512 during encoding and halved from 512 to 64 during decoding.
We set the combination of dice loss $\mathit{L_d}$ and cross-entropy loss $\mathit{L_c}$ as the loss function using the ground-truth label map.
The final loss function is $0.5*\mathit{L_d} + 0.5*\mathit{L_c}$.

\textbf{Training details.} We implement our COVID-SegNet using Pytorch. For network training, we train all models from scratch with random initial parameters. 
The entire models are conducted on a server with six Nvidia TITAN RTX GPUs with 24 GB memory.
We randomly crop the $128 \times 128 \times 64$ patches as the training samples.
For optimization, we use Adam optimizer by setting $\beta_1=0.9$, $\beta_2 = 0.999$, $\epsilon=10^{-8}$ and batch size is 2.
In experiments, the initial learning rate is $1e^{-4}$, and the learning rate decay of $1e^{-6}$.
The proposed network will preform both lung and COVID-19 segmentation tasks.

\def \Lwid{0.32}
\def \patchwid{0.075}
\def \Imagewidth{1.575in}
\def \patchwidth{0.43in}

\fboxsep=0mm
\fboxrule=0.9pt
\def \pOneColor{red}
\def \pTwoColor{blue}
\def \pThreeColor{cyan}

\def \pOneColorShade{red}
\def \pTwoColorShade{red}
\def \pThreeColorShade{red}
\definecolor{ColorName}{RGB}{34,251,250}

\begin{figure*}[htbp]
\centering

\begin{minipage}[h]{\Lwid\linewidth}
\centering
\includegraphics[trim=200 10 100 60, clip, height=\Imagewidth]{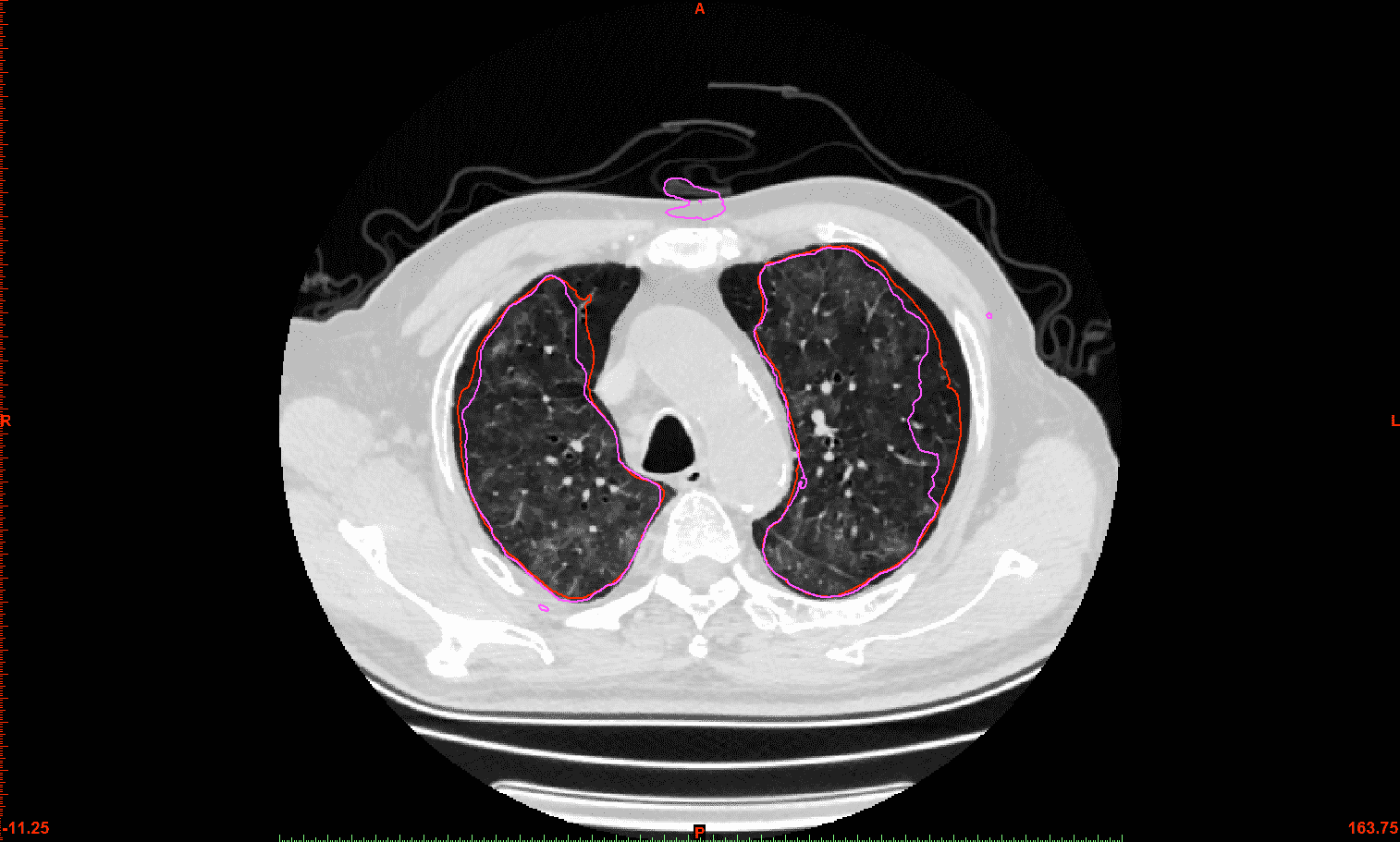}
\centerline{\small (a) FCN \cite{yang2019fd}}
\end{minipage}%
\begin{minipage}[h]{\Lwid\linewidth}
\centering
\includegraphics[trim=200 10 100 60, clip, height=\Imagewidth]{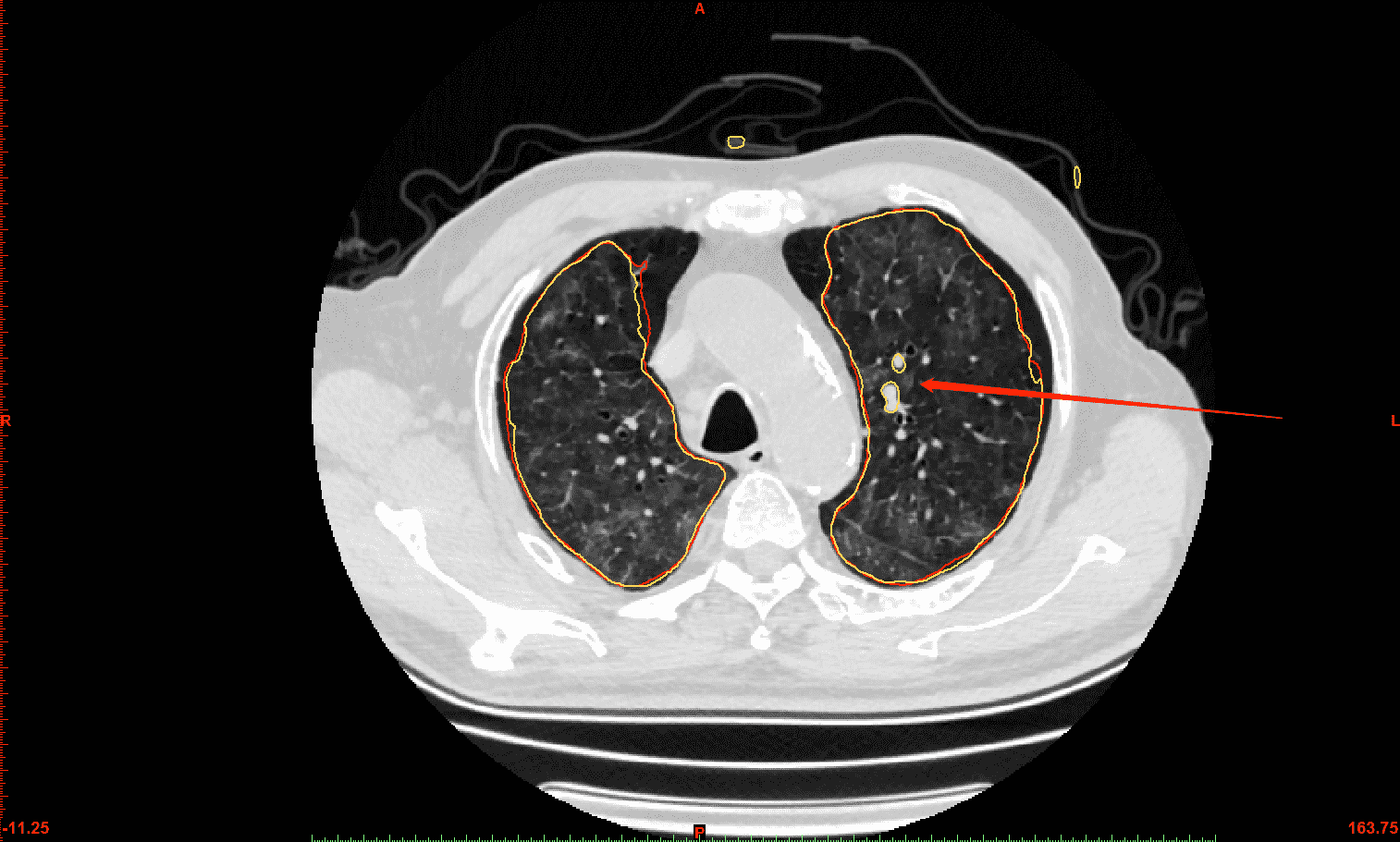}
\centerline{\small (b) UNet \cite{cciccek20163d}}
\end{minipage}%
\begin{minipage}[h]{\Lwid\linewidth}
\centering
\includegraphics[trim=200 10 100 60, clip, height=\Imagewidth]{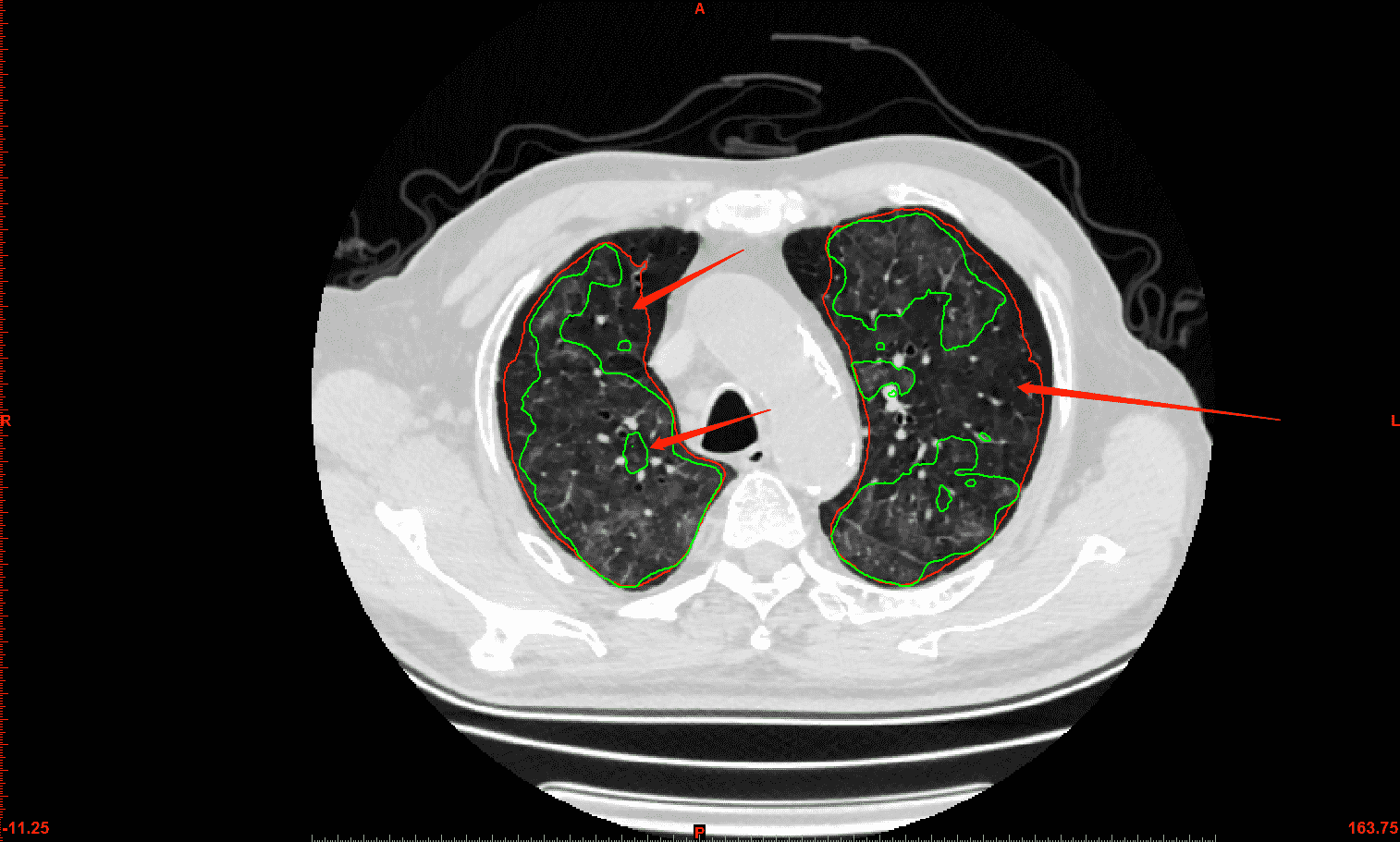}
\centerline{\small (c) UNet++ \cite{zhou18deep}}
\end{minipage}%

\begin{minipage}[h]{\Lwid\linewidth}
\centering
\includegraphics[trim=200 10 100 60, clip, height=\Imagewidth]{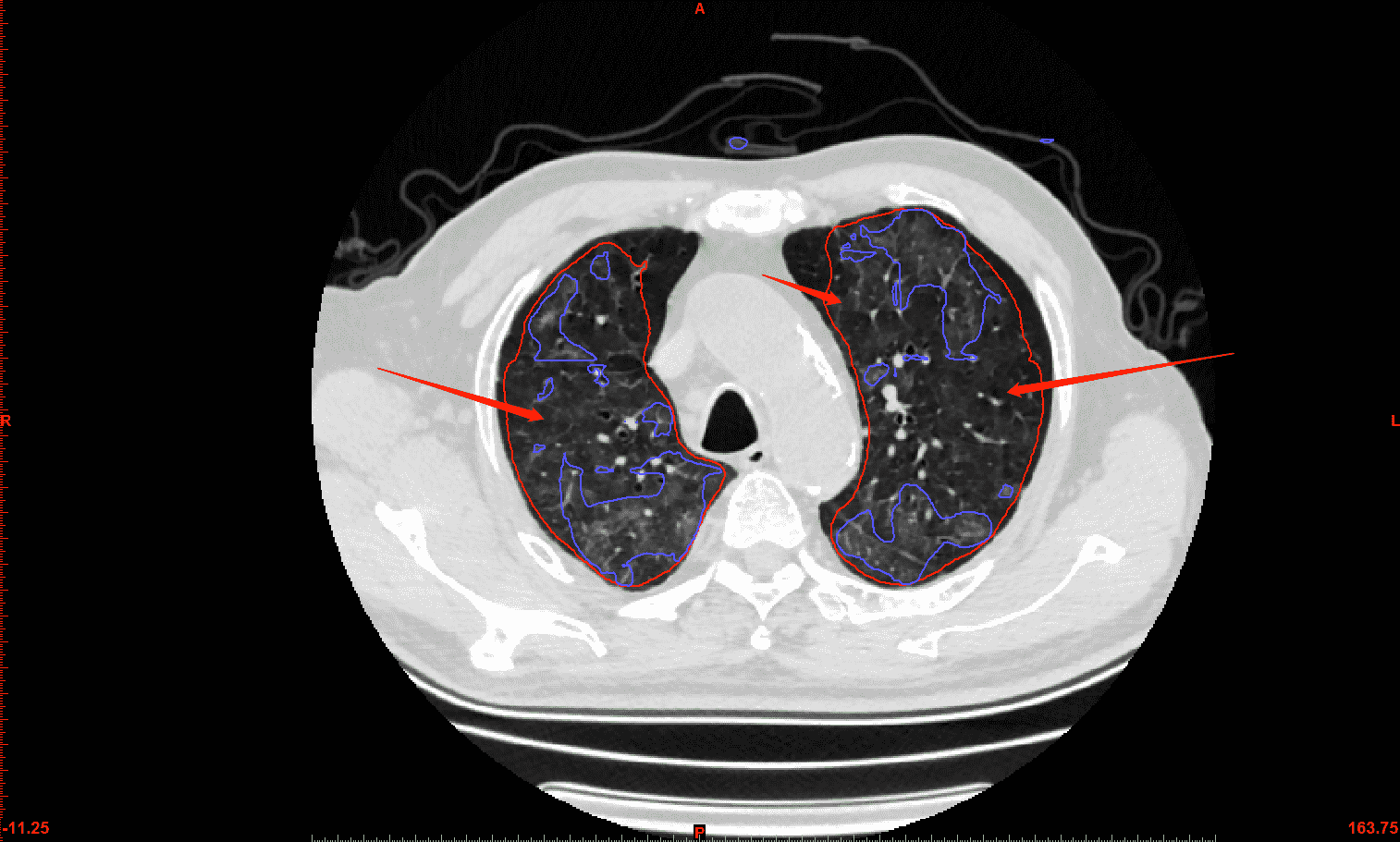}
\centerline{\small (d) VNet \cite{milletari2016v}}
\end{minipage}%
\begin{minipage}[h]{\Lwid\linewidth}
\centering
\includegraphics[trim=200 10 100 60, clip, height=\Imagewidth]{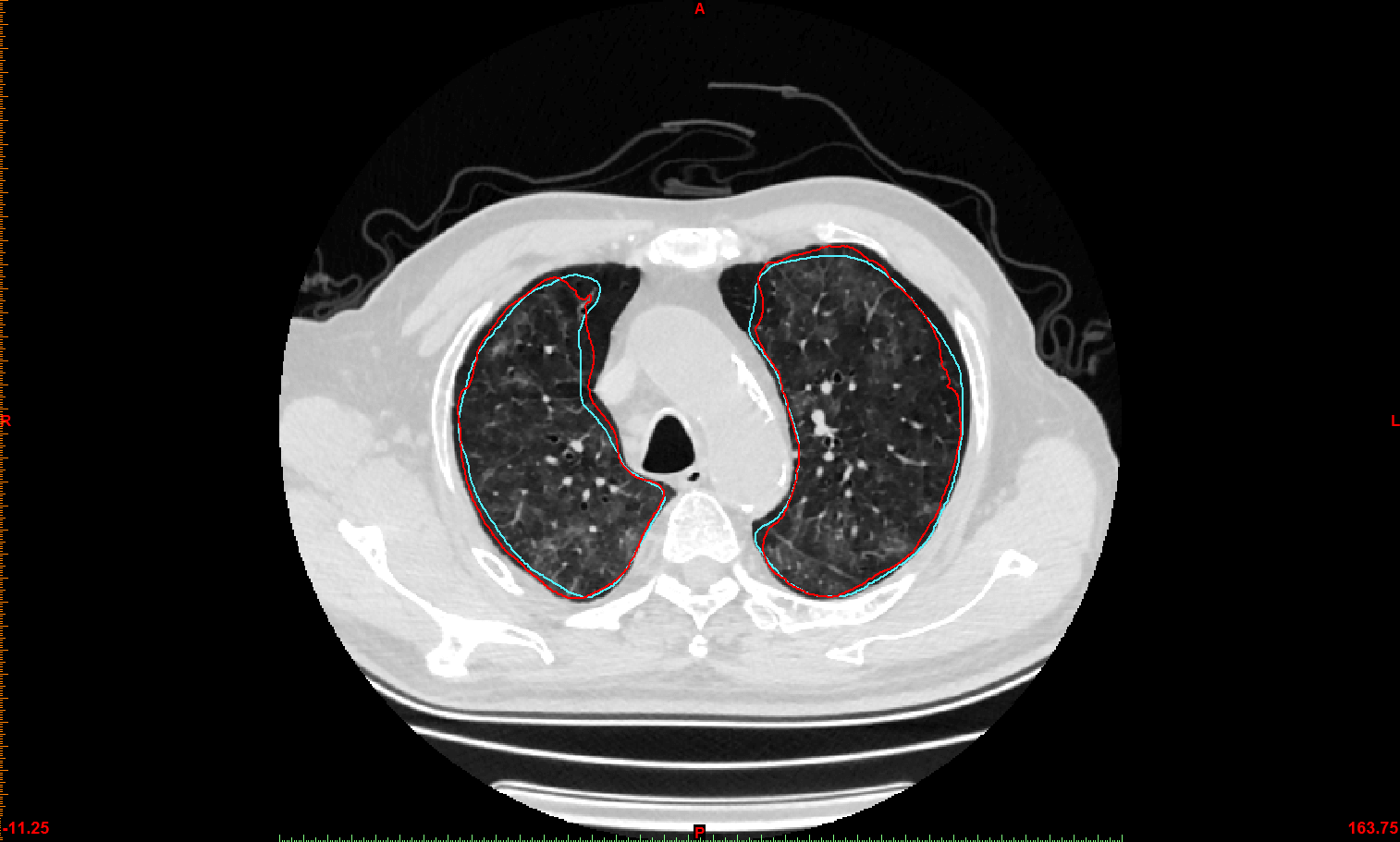}
\centerline{\small (e) Ours }
\end{minipage}%
\begin{minipage}[h]{\Lwid\linewidth}
\centering
\includegraphics[trim=200 10 100 60, clip, height=\Imagewidth]{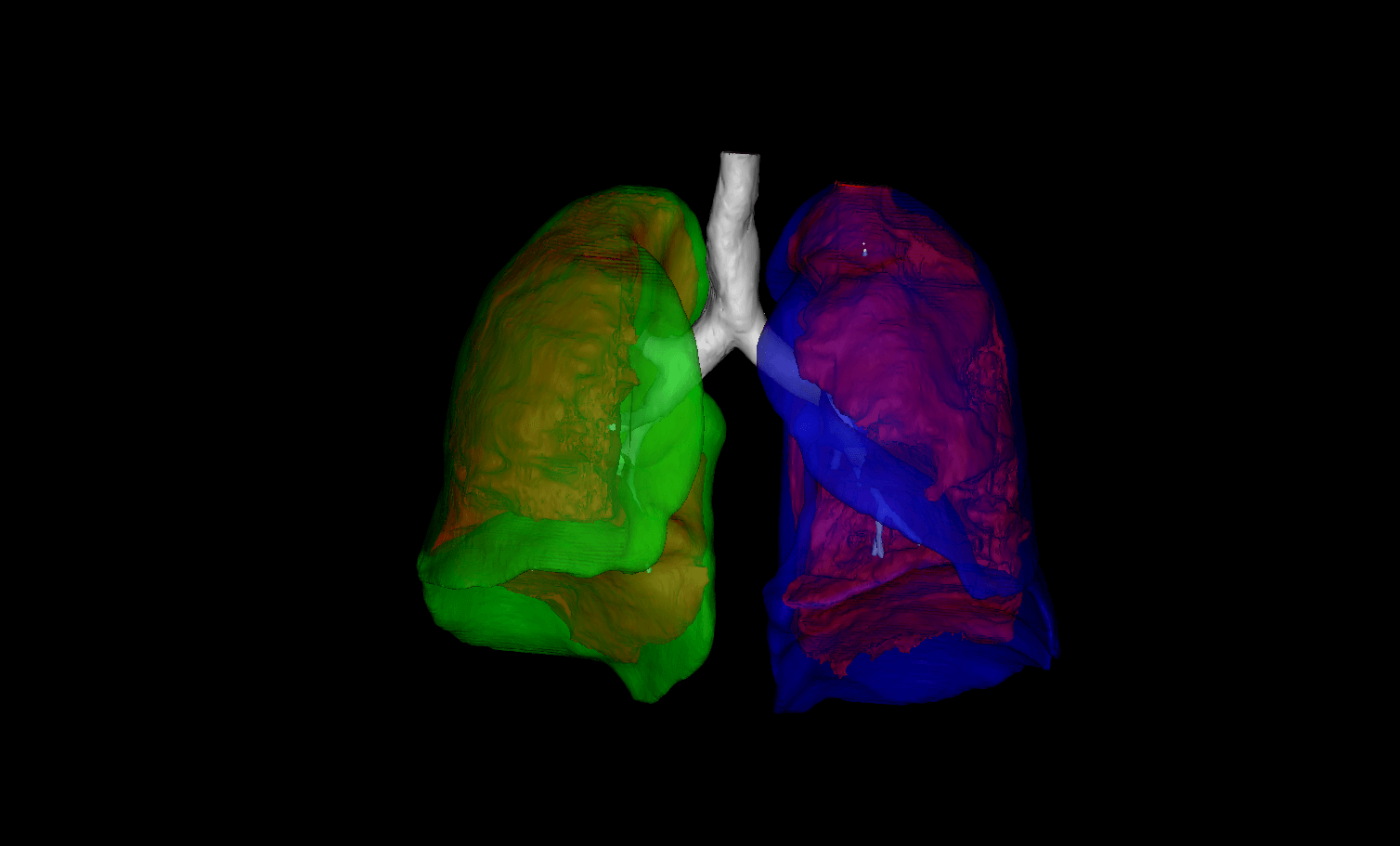}
\centerline{\small (f) Surface rendering of ours }
\end{minipage}%

\caption{Visual comparisons on the testing data for COVID-19 segmentation.
(a)-(e) show the results of the state-of-the-art methods and the proposed method, respectively.
(f) is the 3D surface rendering of COVD-19 infections (severe) segmented by our method.
The red arrows indicate the flows of different methods. Ground truth is shown with the red line, other methods are displayed with different colors.}
\label{fig_test_1}
\end{figure*}

\begin{figure*}[h!]
\centering

\begin{minipage}[h]{\Lwid\linewidth}
\centering
\includegraphics[trim=200 10 100 60, clip, height=\Imagewidth]{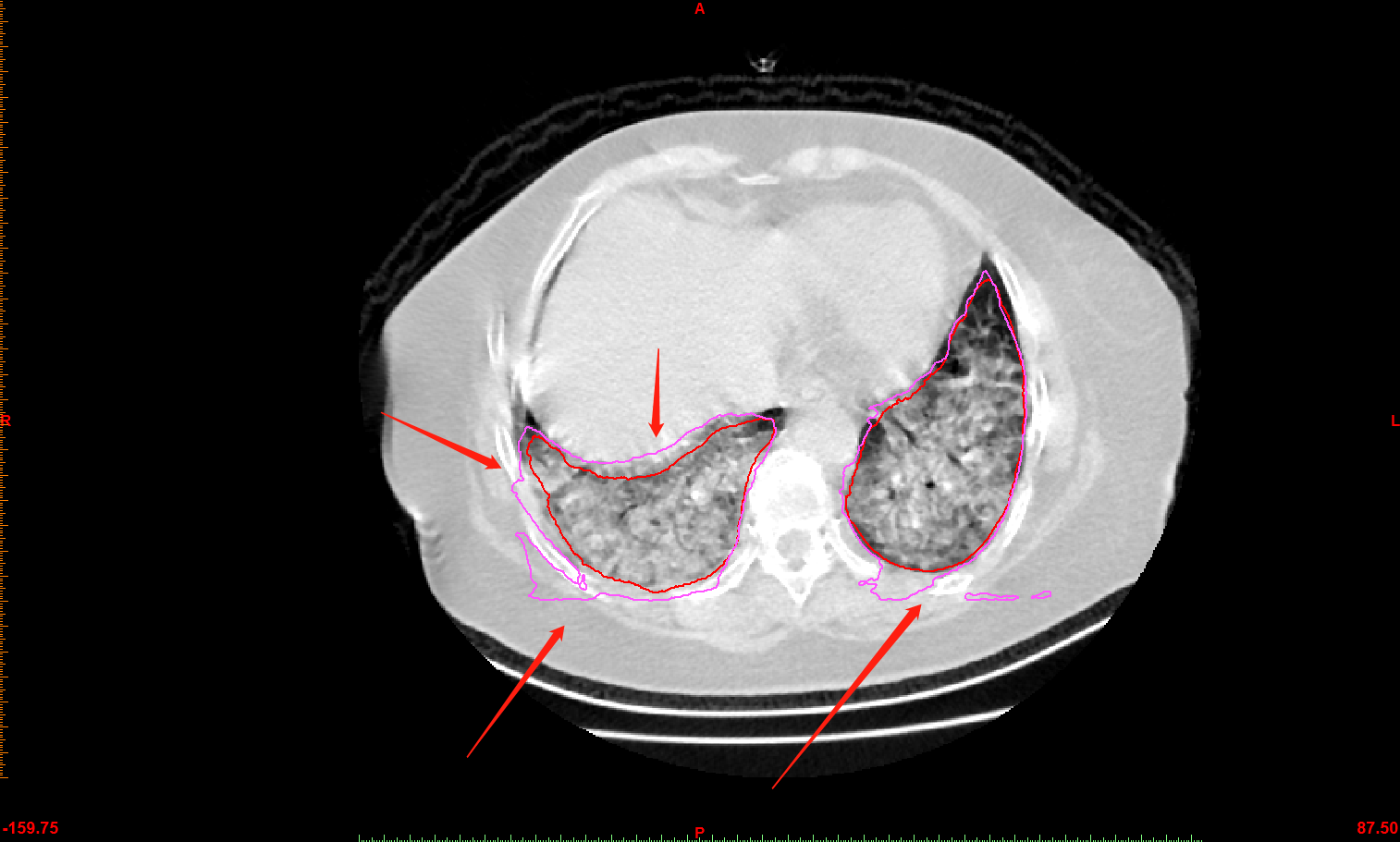}
\centerline{\small (a) FCN \cite{yang2019fd}}
\end{minipage}%
\begin{minipage}[h]{\Lwid\linewidth}
\centering
\includegraphics[trim=200 10 100 60, clip, height=\Imagewidth]{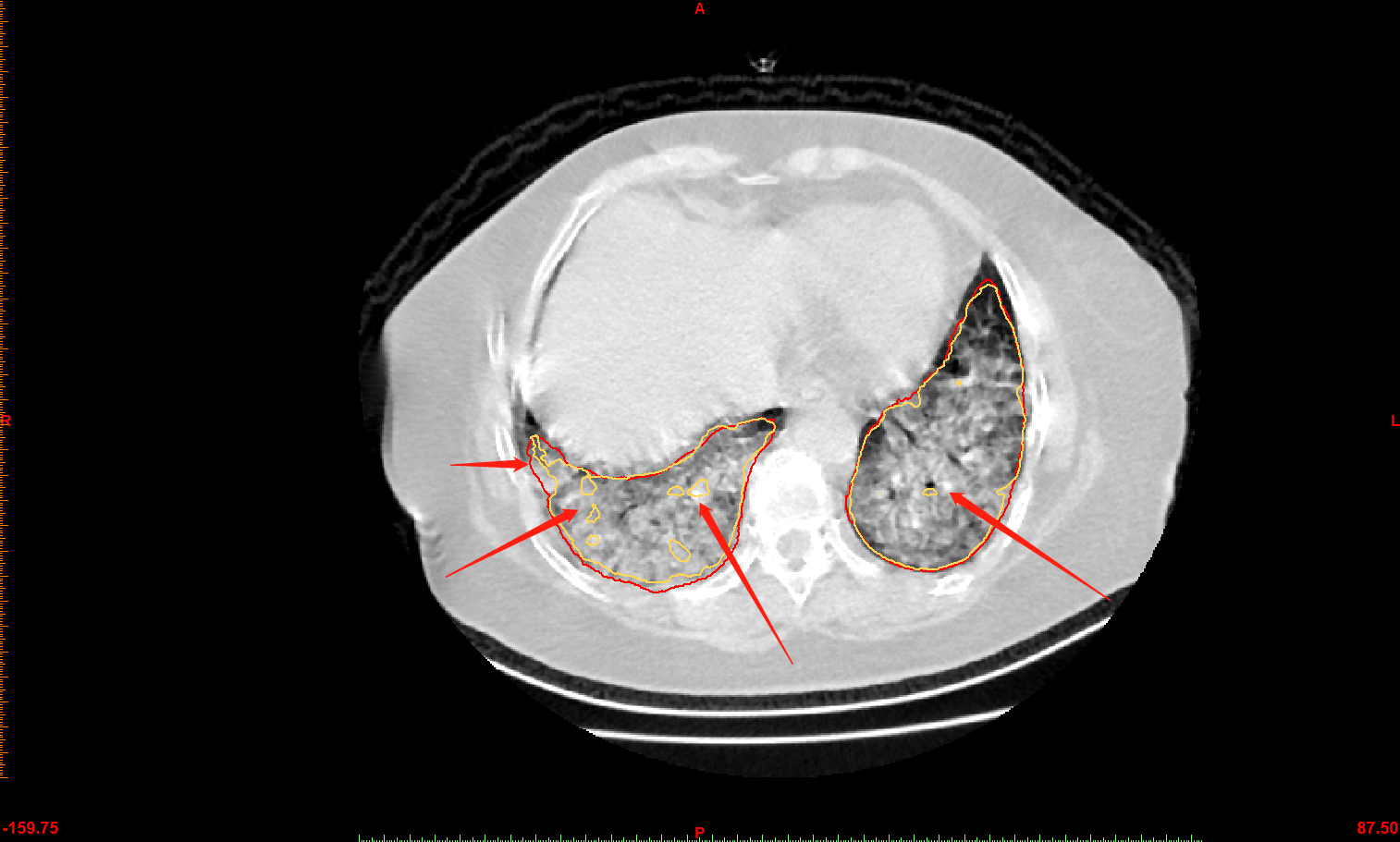}
\centerline{\small (b) UNet \cite{cciccek20163d}}
\end{minipage}%
\begin{minipage}[h]{\Lwid\linewidth}
\centering
\includegraphics[trim=200 10 100 60, clip, height=\Imagewidth]{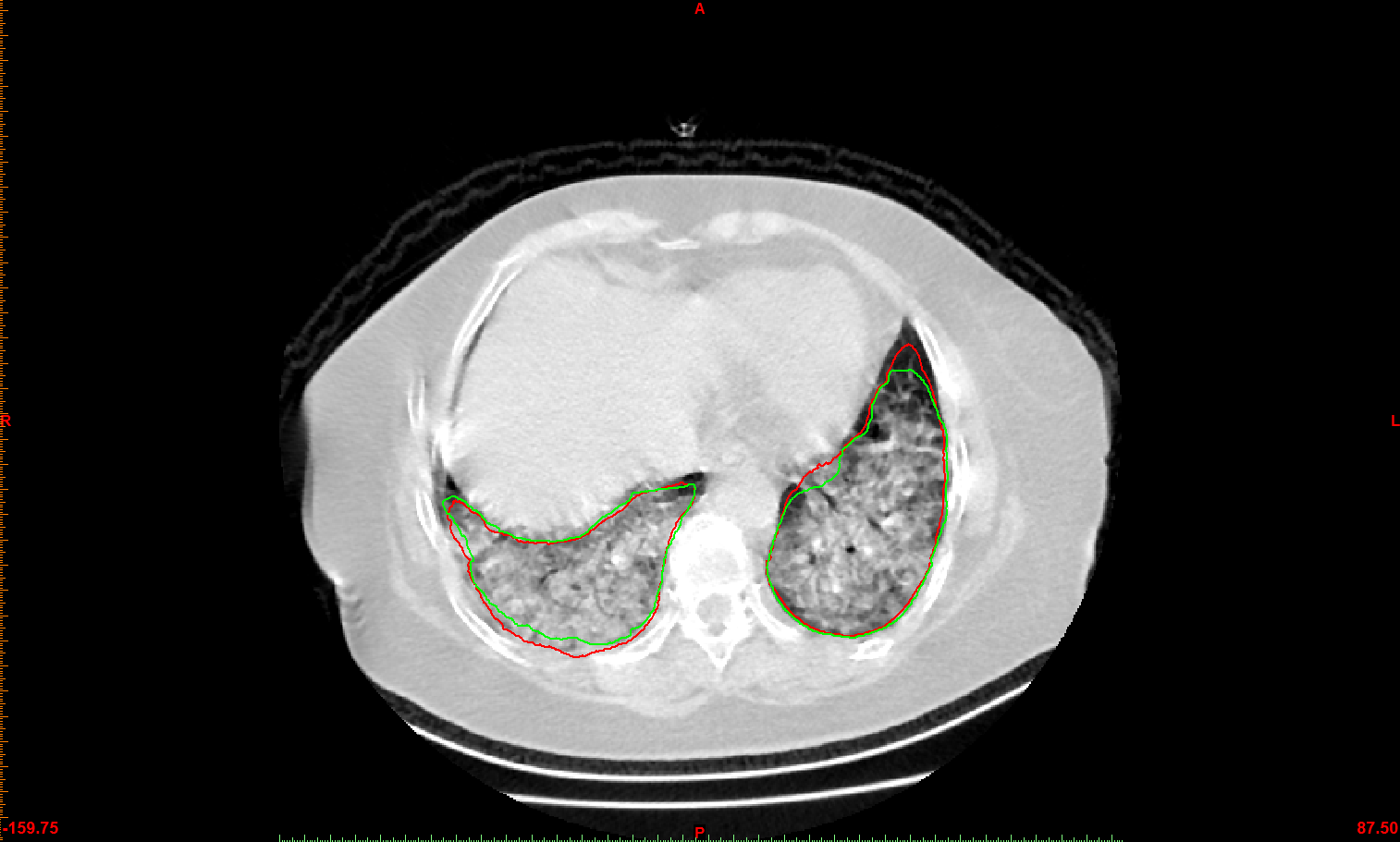}
\centerline{\small (c) UNet++ \cite{zhou18deep}
}
\end{minipage}%

\begin{minipage}[h]{\Lwid\linewidth}
\centering
\includegraphics[trim=200 10 100 60, clip, height=\Imagewidth]{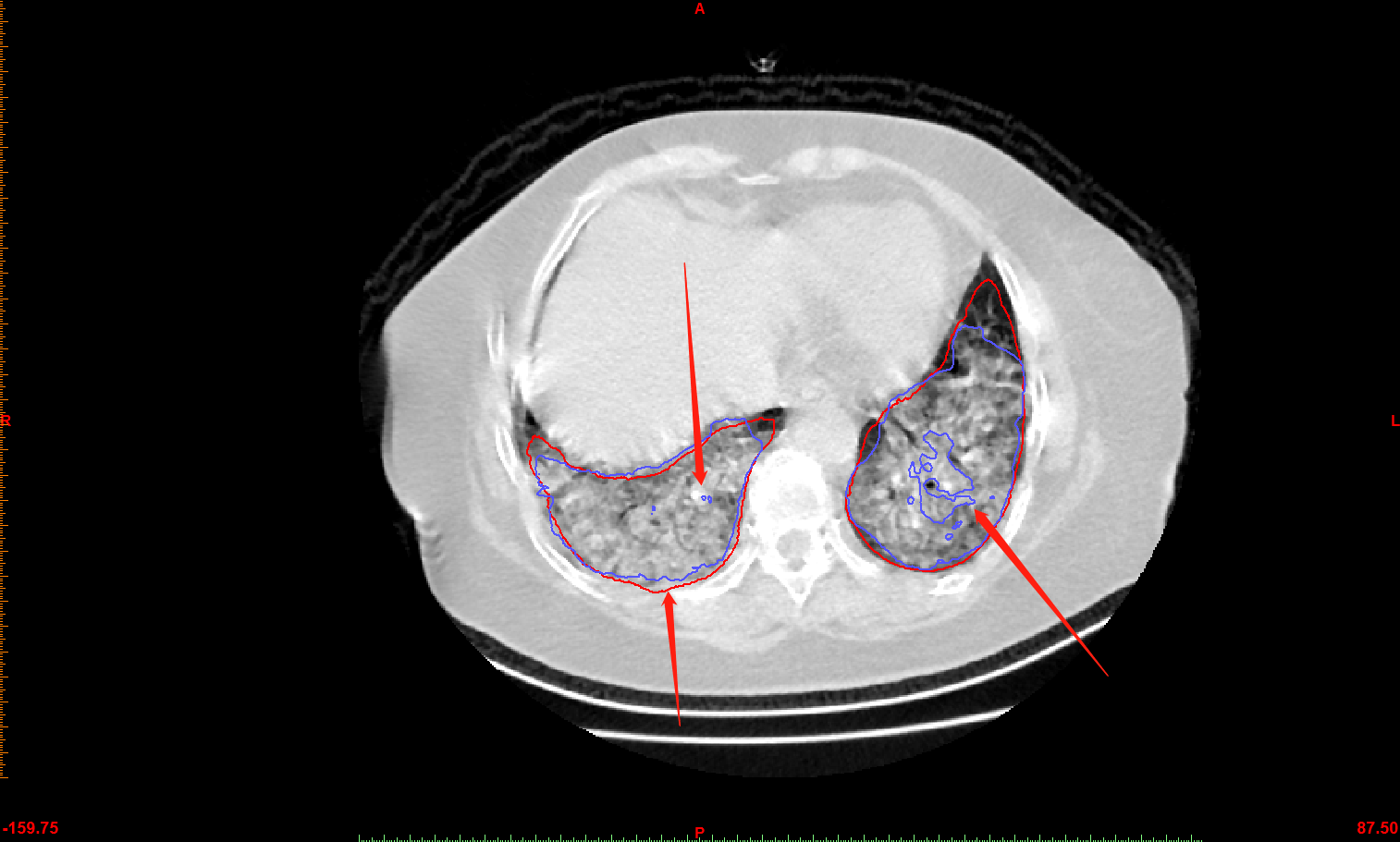}
\centerline{\small (d) VNet \cite{milletari2016v} }
\end{minipage}%
\begin{minipage}[h]{\Lwid\linewidth}
\centering
\includegraphics[trim=200 10 100 60, clip, height=\Imagewidth]{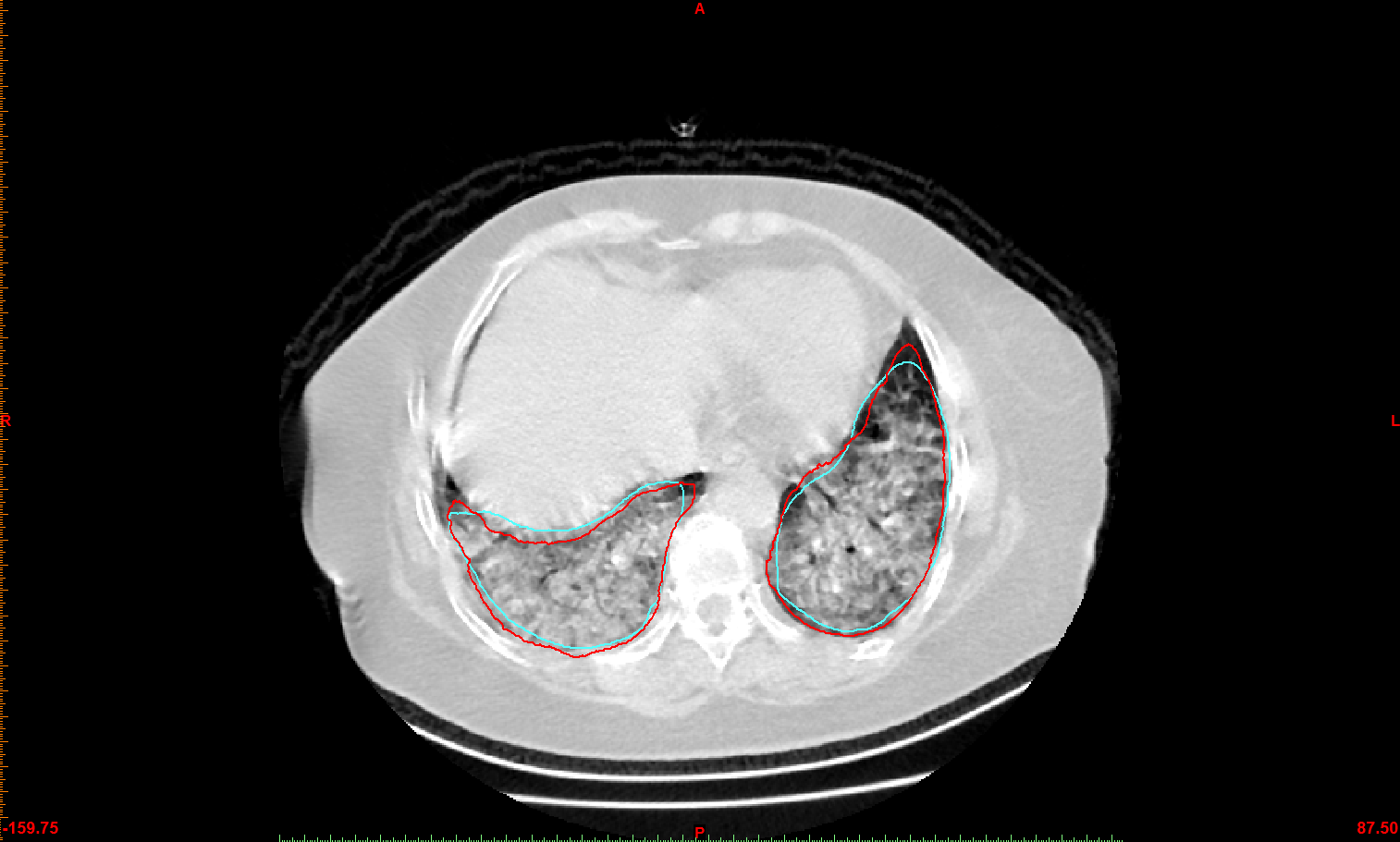}
\centerline{\small (e) Ours }
\end{minipage}%
\begin{minipage}[h]{\Lwid\linewidth}
\centering
\includegraphics[trim=200 10 100 60, clip, height=\Imagewidth]{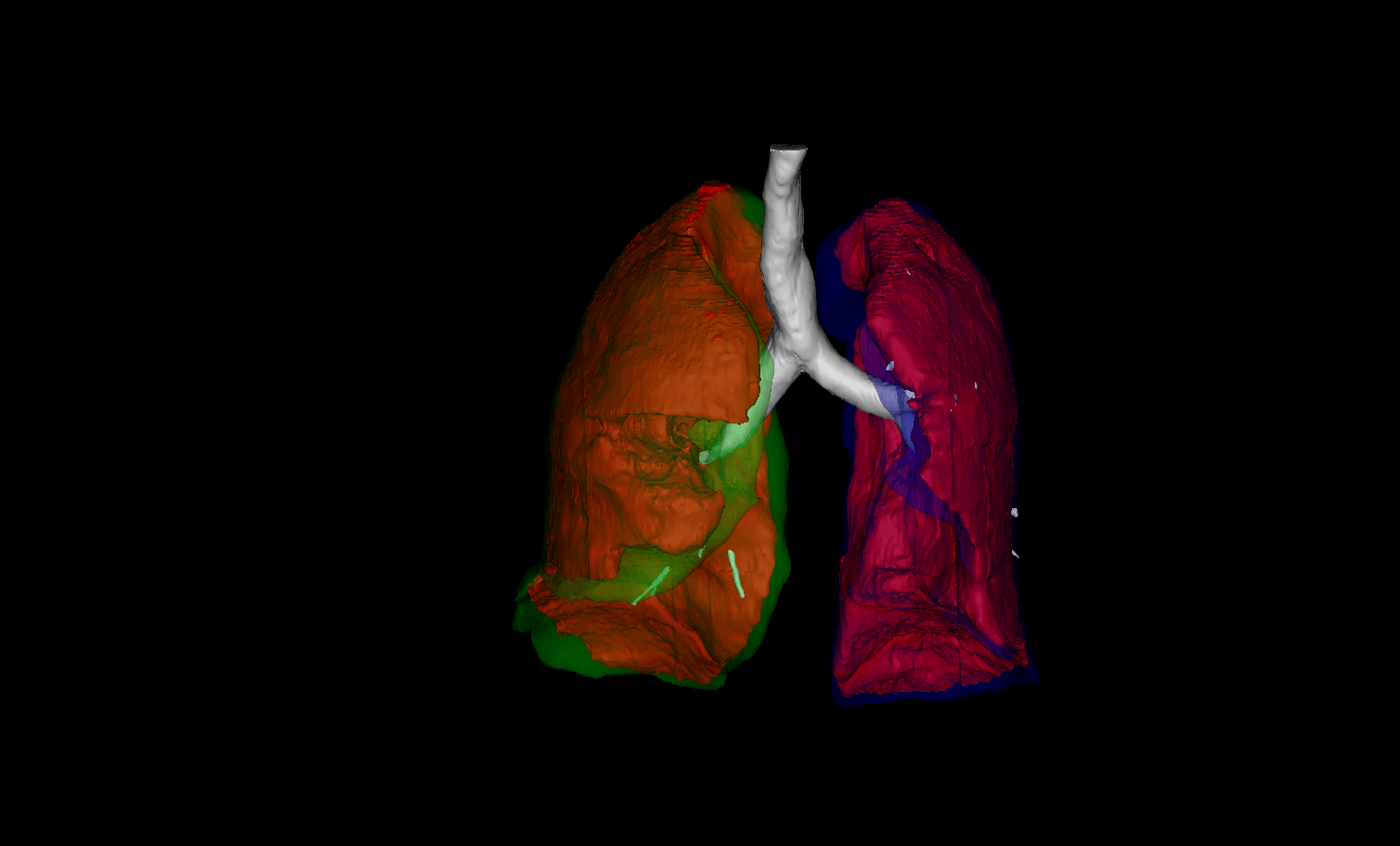}
\centerline{\small (f) Surface rendering of ours }
\end{minipage}%

\caption{Typical infection segmentation results of CT scans of COVID-19 patient (severe). The contrast of this case is too low to segment COVID-19 infection. The proposed method still can handle this difficulty sample. The red arrows indicate the flows of different methods. Ground truth is shown with the red line, other methods are displayed with different colors.}
\label{fig_test_2}
\end{figure*}

\begin{figure*}[h]
\centering

\begin{minipage}[h]{\Lwid\linewidth}
\centering
\includegraphics[trim=200 10 100 60, clip, height=\Imagewidth]{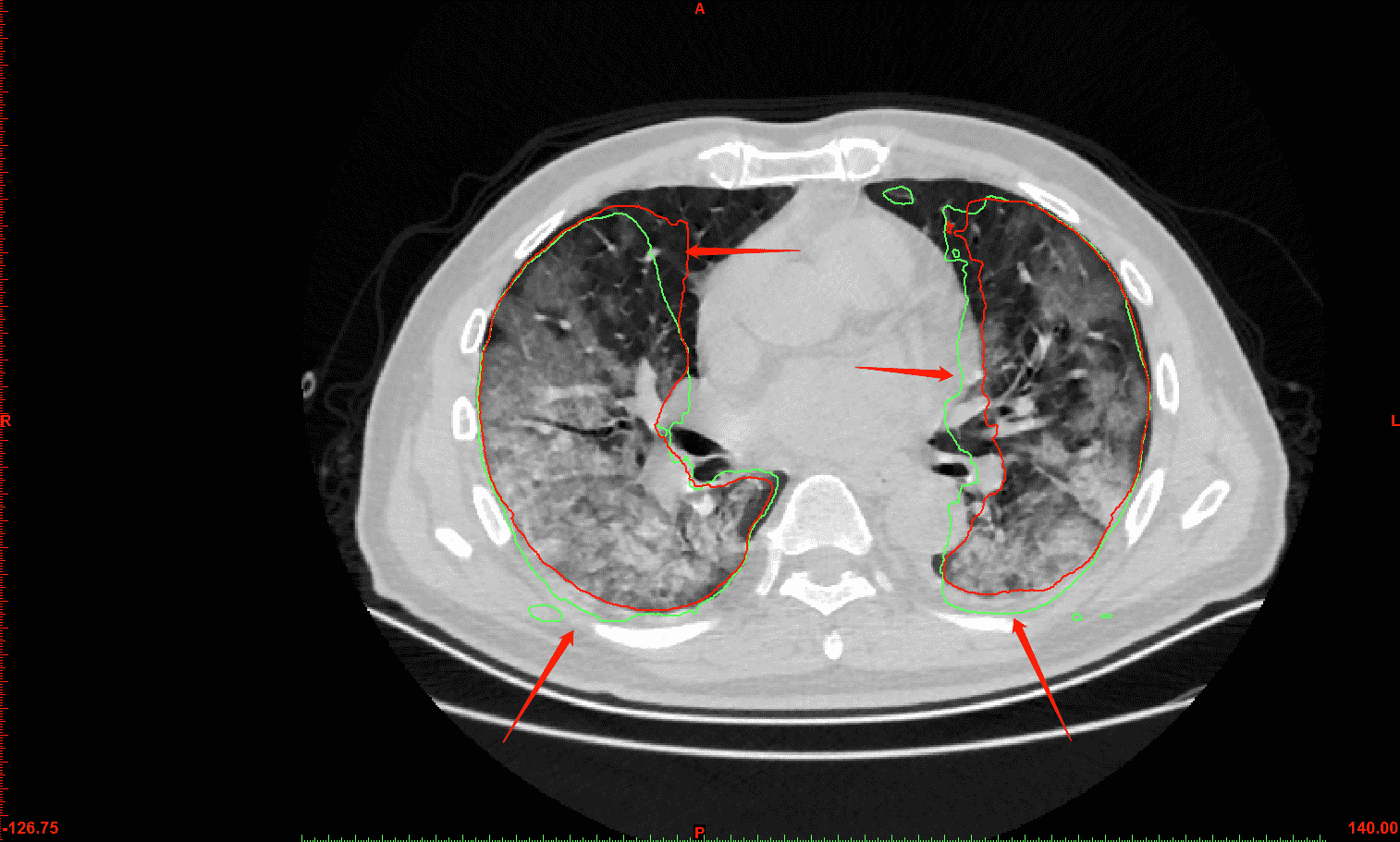}
\centerline{\small (a) FCN \cite{yang2019fd}}
\end{minipage}%
\begin{minipage}[h]{\Lwid\linewidth}
\centering
\includegraphics[trim=200 10 100 60, clip, height=\Imagewidth]{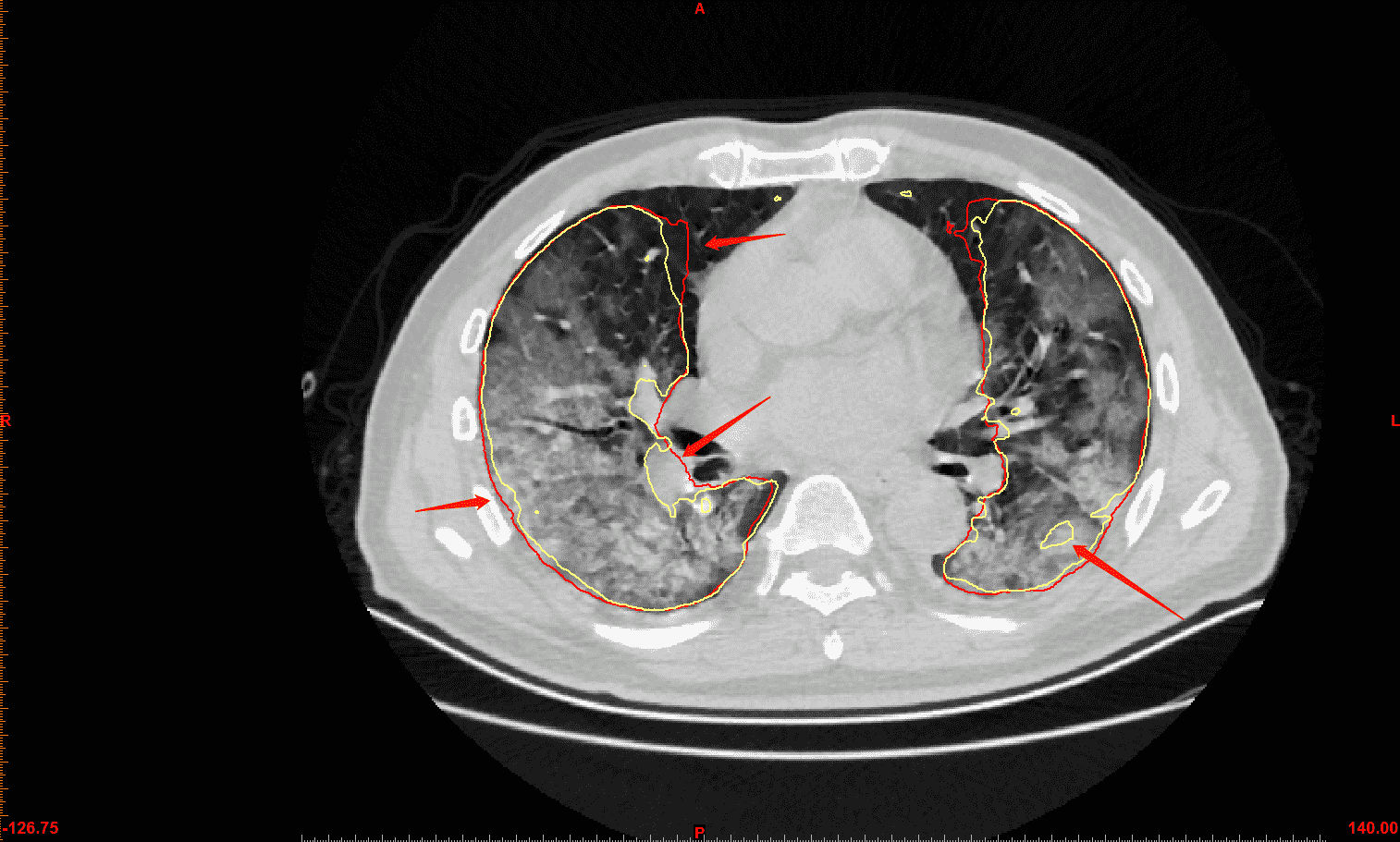}
\centerline{\small (b) UNet \cite{cciccek20163d}}
\end{minipage}%
\begin{minipage}[h]{\Lwid\linewidth}
\centering
\includegraphics[trim=200 10 100 60, clip, height=\Imagewidth]{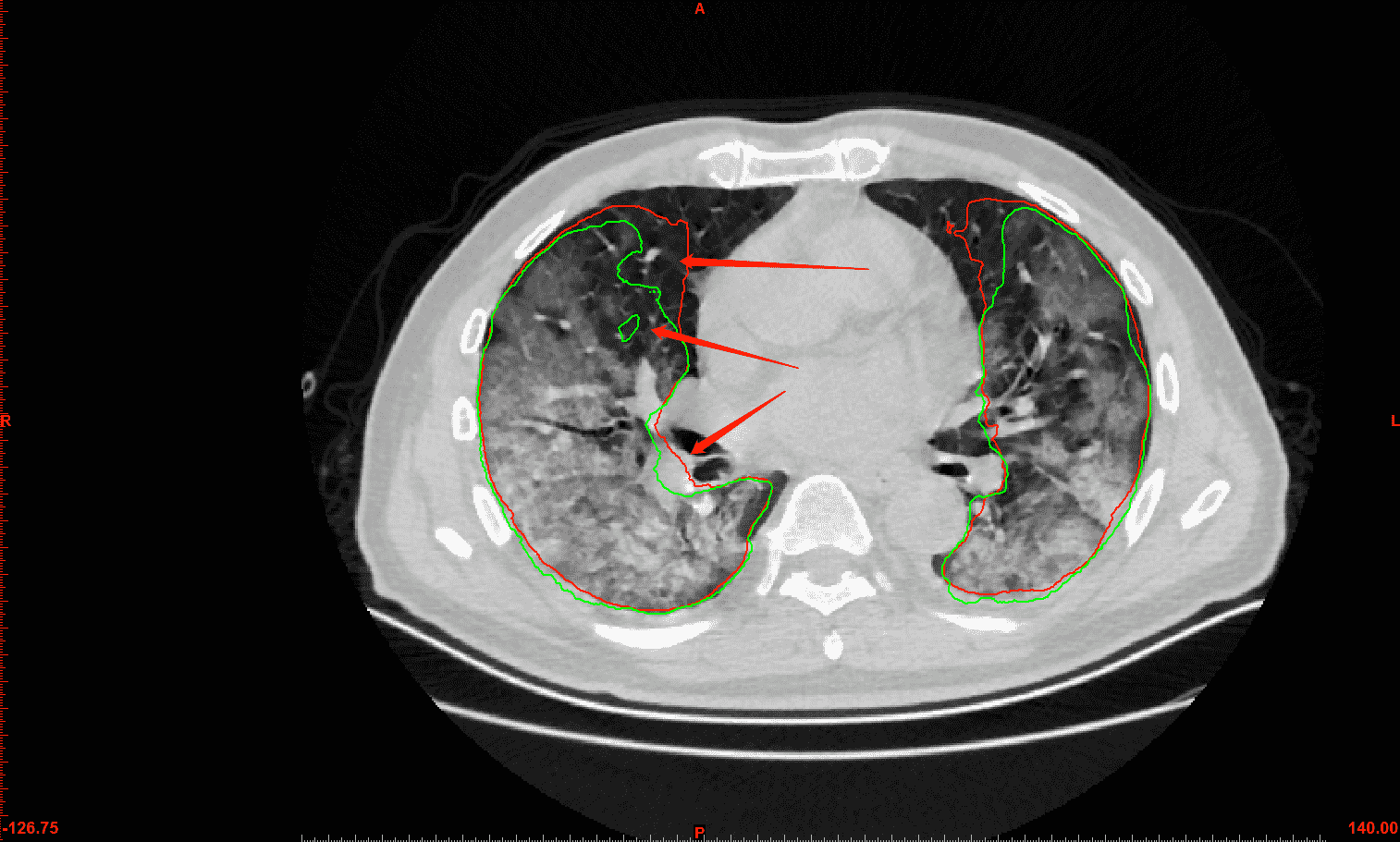}
\centerline{\small (c) UNet++ \cite{zhou18deep}}
\end{minipage}%

\begin{minipage}[h]{\Lwid\linewidth}
\centering
\includegraphics[trim=200 10 100 60, clip, height=\Imagewidth]{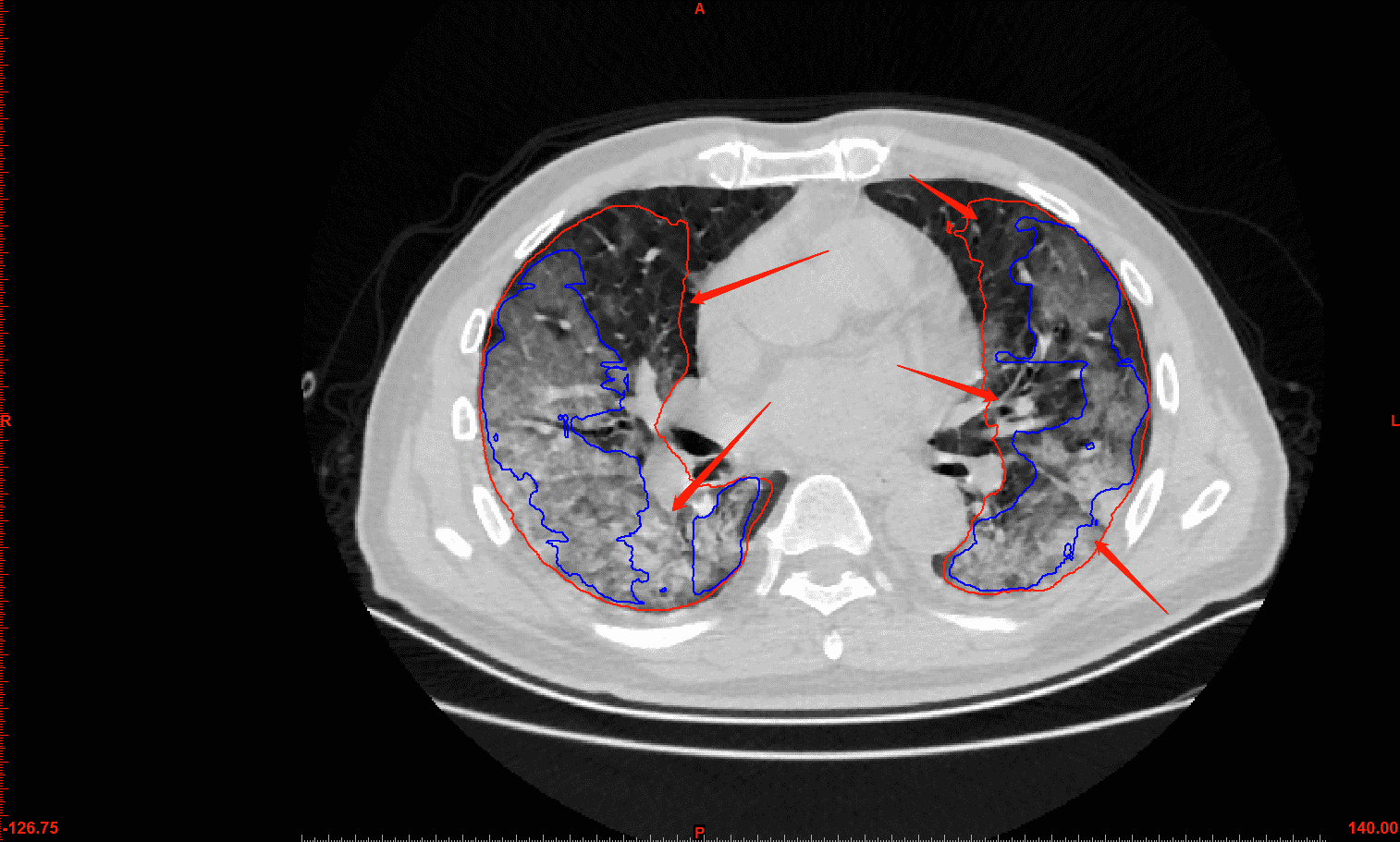}
\centerline{\small (d) VNet \cite{milletari2016v} }
\end{minipage}%
\begin{minipage}[h]{\Lwid\linewidth}
\centering
\includegraphics[trim=200 10 100 60, clip, height=\Imagewidth]{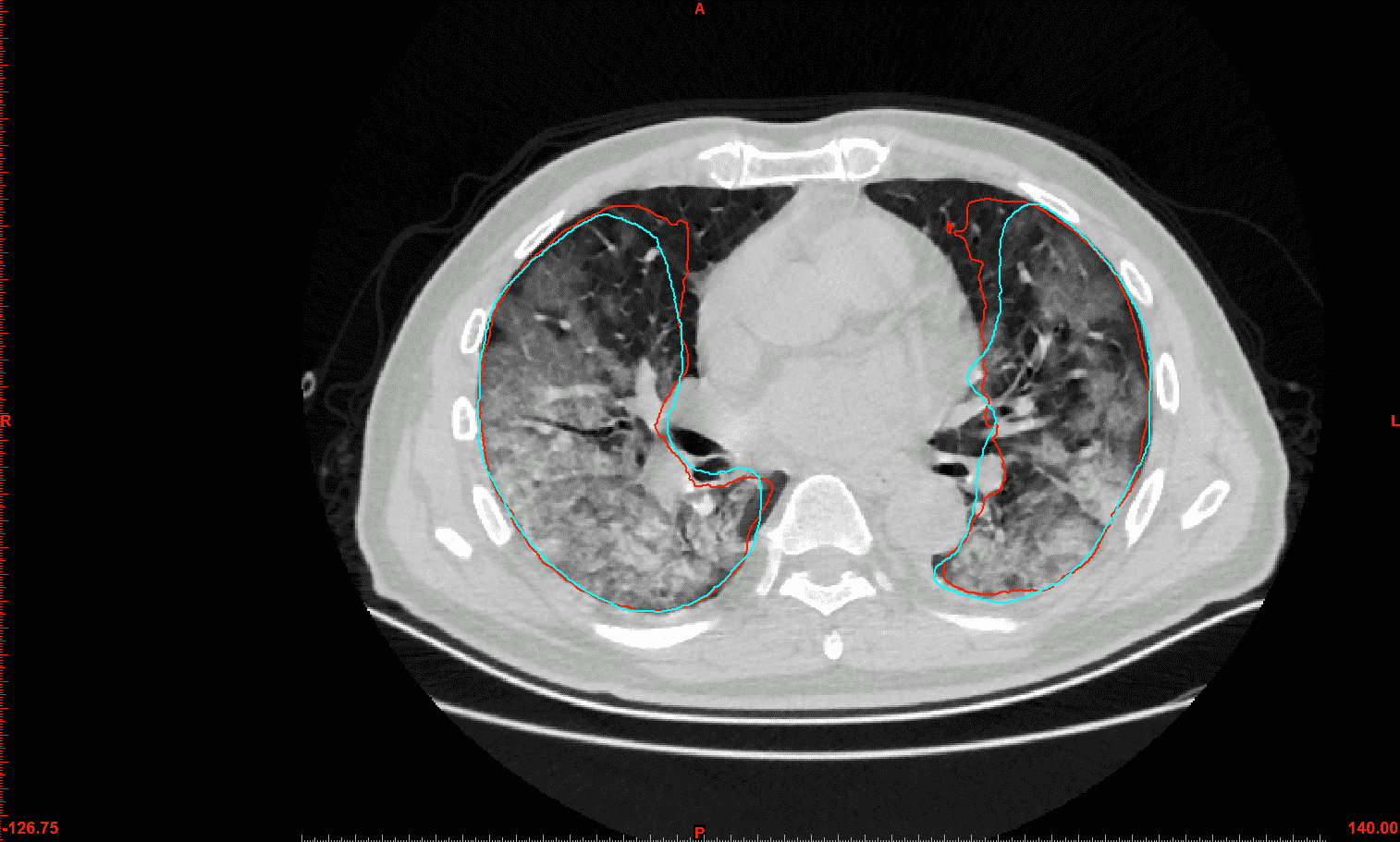}
\centerline{\small (e) Ours }
\end{minipage}%
\begin{minipage}[h]{\Lwid\linewidth}
\centering
\includegraphics[trim=200 10 100 60, clip, height=\Imagewidth]{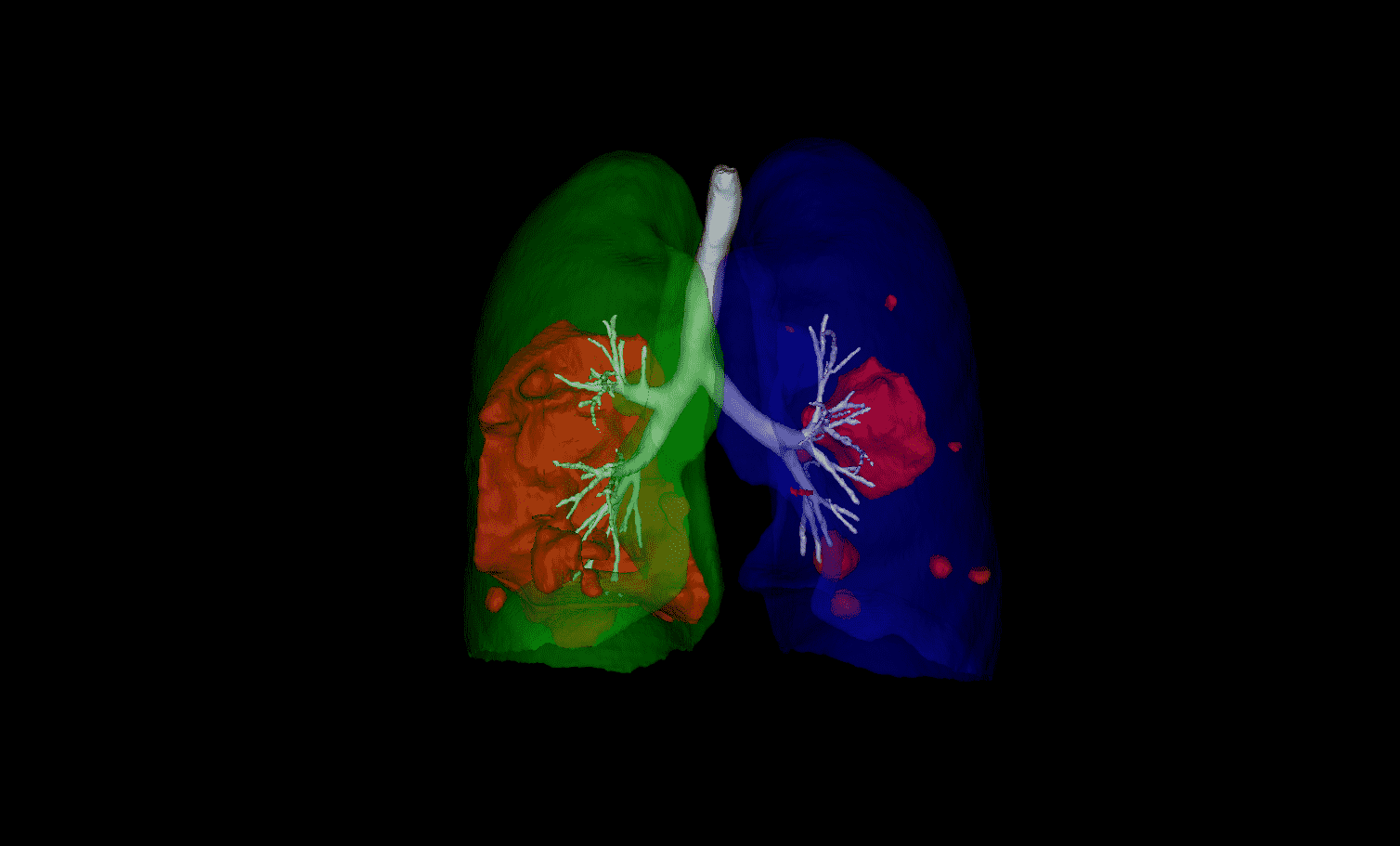}
\centerline{\small (f) Surface rendering of ours }
\end{minipage}%

\caption{Comparisons on the chest CT example of non-severe infection COVID-19 on the test set. The infection regions are not easily to peeling from the chest wall. The red arrows indicate the flows of different methods. Ground truth is shown with the red line, other methods are displayed with different colors.}
\label{fig_test_3}
\end{figure*}

\begin{figure*}[h!]
\centering

\begin{minipage}[h]{\Lwid\linewidth}
\centering
\includegraphics[trim=200 10 100 60, clip, height=\Imagewidth]{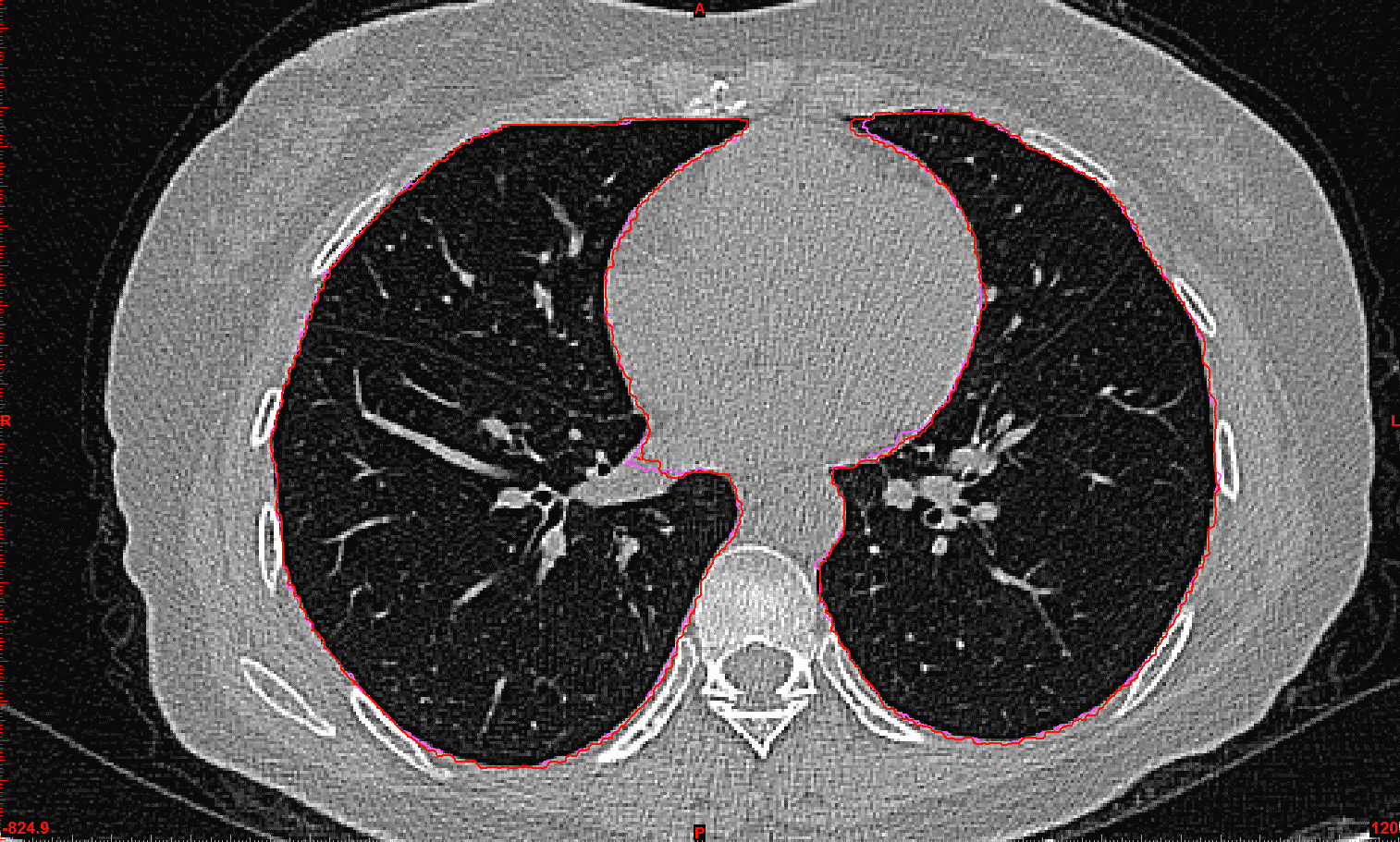}
\centerline{\small (a) FCN \cite{yang2019fd}}
\end{minipage}%
\begin{minipage}[h]{\Lwid\linewidth}
\centering
\includegraphics[trim=200 10 100 60, clip, height=\Imagewidth]{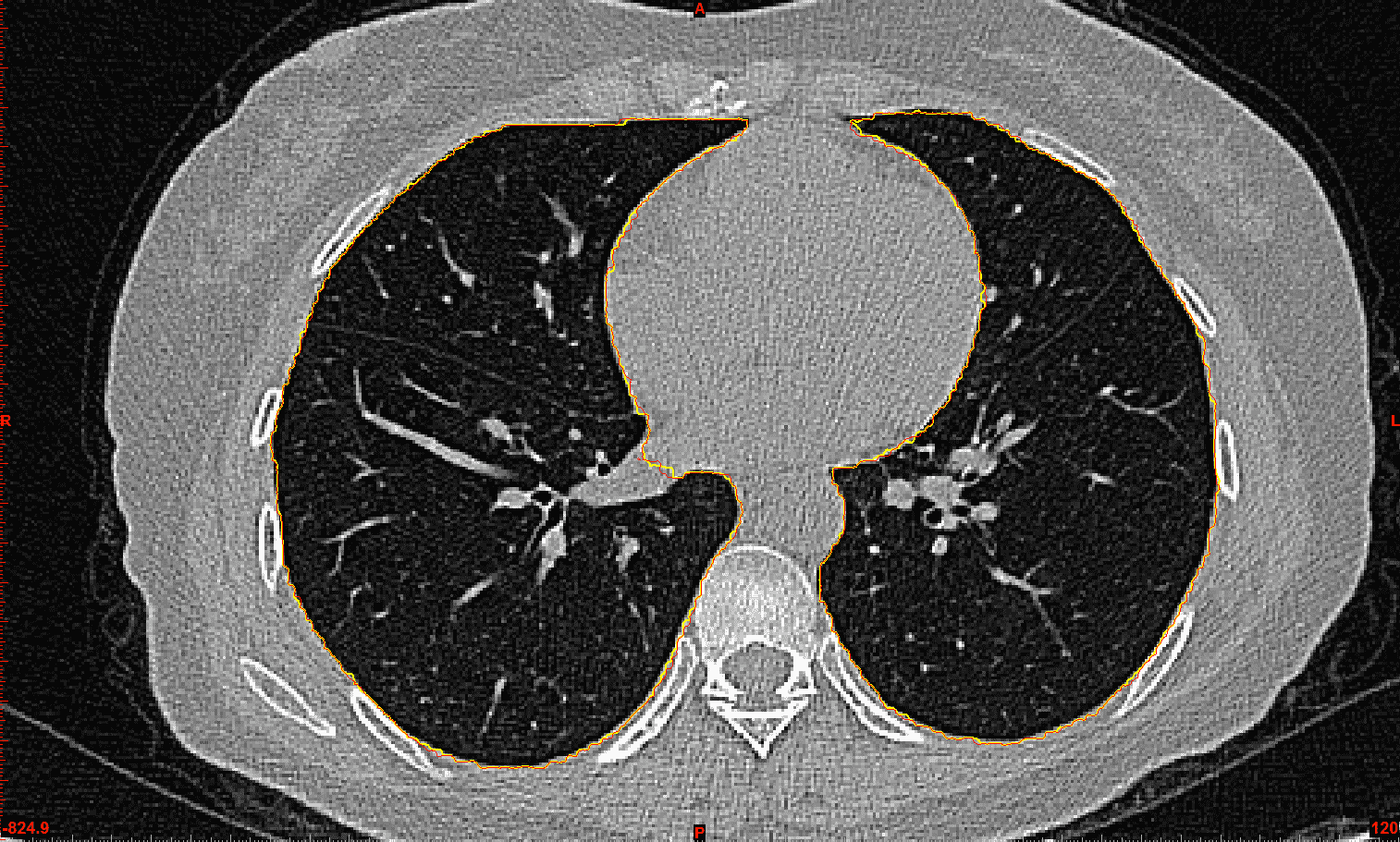}
\centerline{\small (b) UNet \cite{cciccek20163d}}
\end{minipage}%
\begin{minipage}[h]{\Lwid\linewidth}
\centering
\includegraphics[trim=200 10 100 60, clip, height=\Imagewidth]{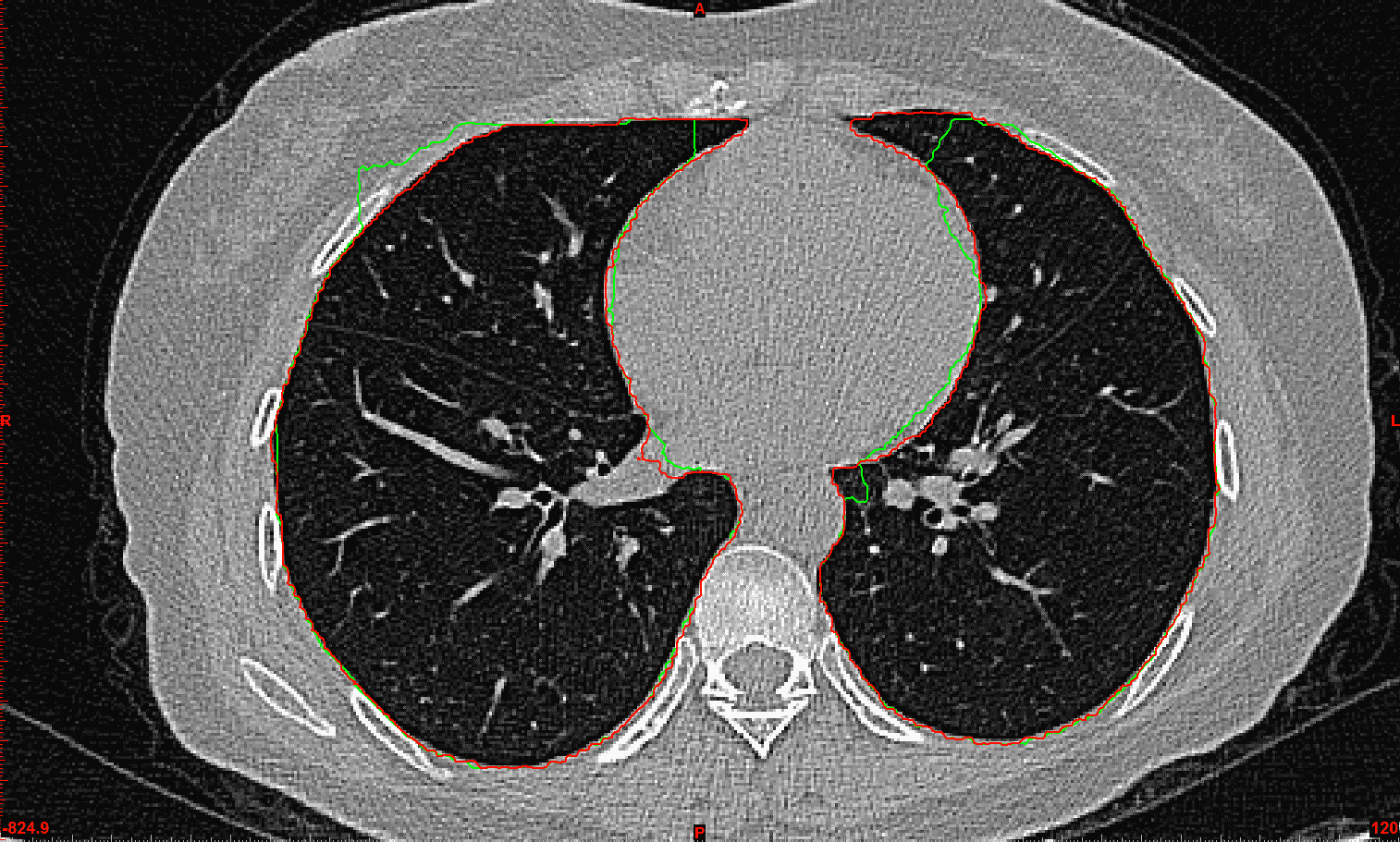}
\centerline{\small (c) UNet++ \cite{zhou18deep}
}
\end{minipage}%

\begin{minipage}[h]{\Lwid\linewidth}
\centering
\includegraphics[trim=200 10 100 60, clip, height=\Imagewidth]{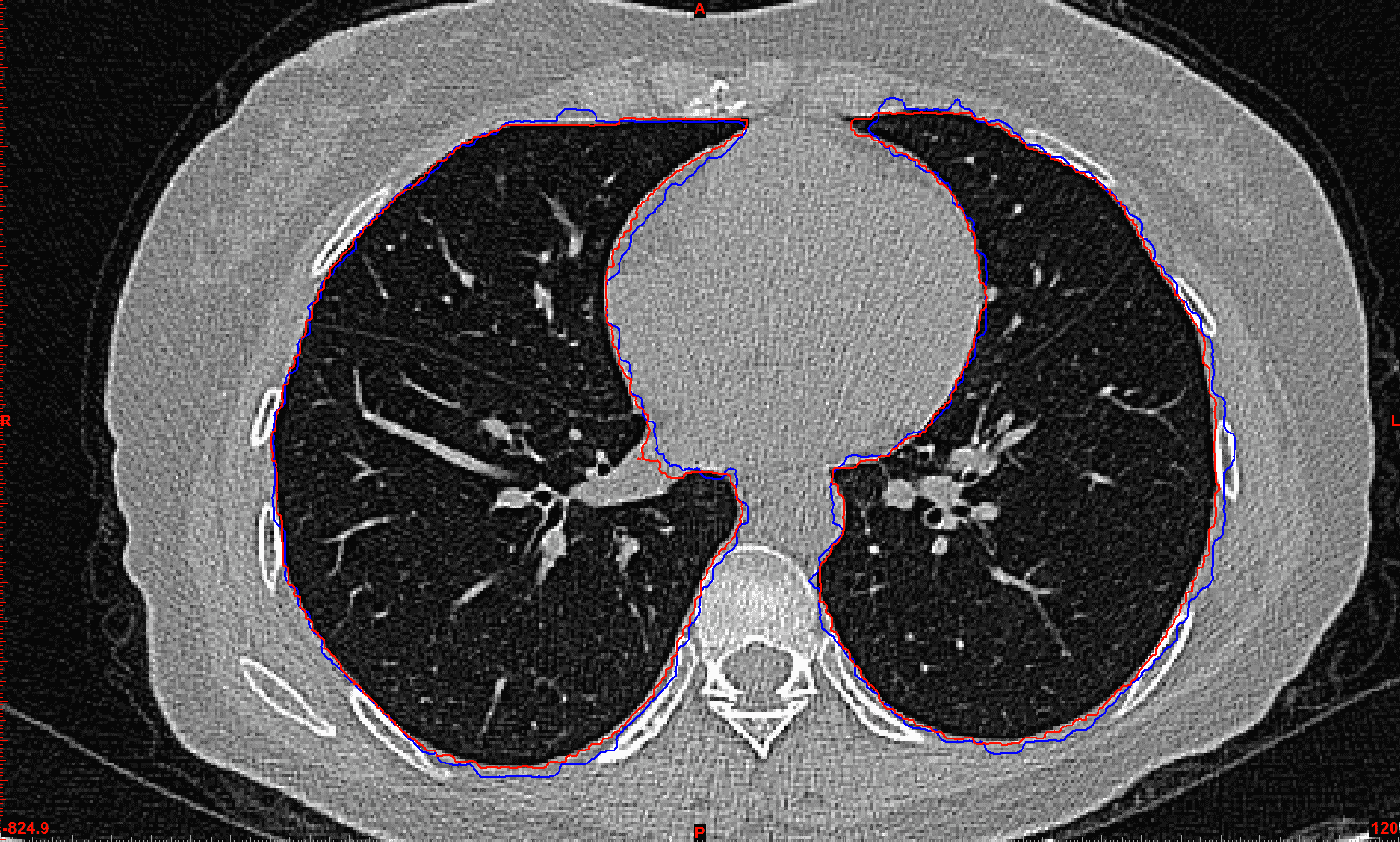}
\centerline{\small (d) VNet \cite{milletari2016v} }
\end{minipage}%
\begin{minipage}[h]{\Lwid\linewidth}
\centering
\includegraphics[trim=200 10 100 60, clip, height=\Imagewidth]{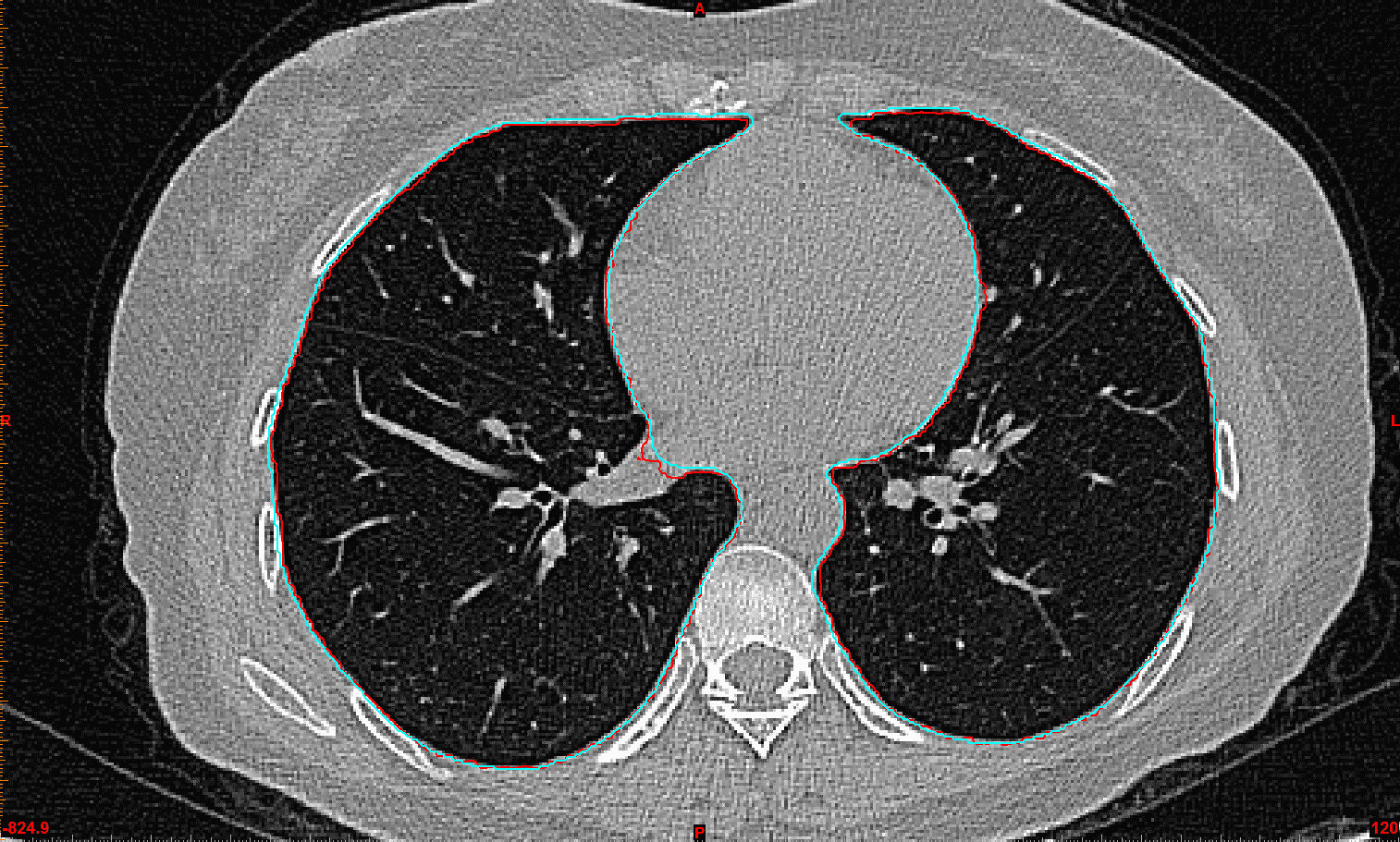}
\centerline{\small (e) Ours }
\end{minipage}%
\begin{minipage}[h]{\Lwid\linewidth}
\centering
\includegraphics[trim=200 10 100 60, clip, height=\Imagewidth]{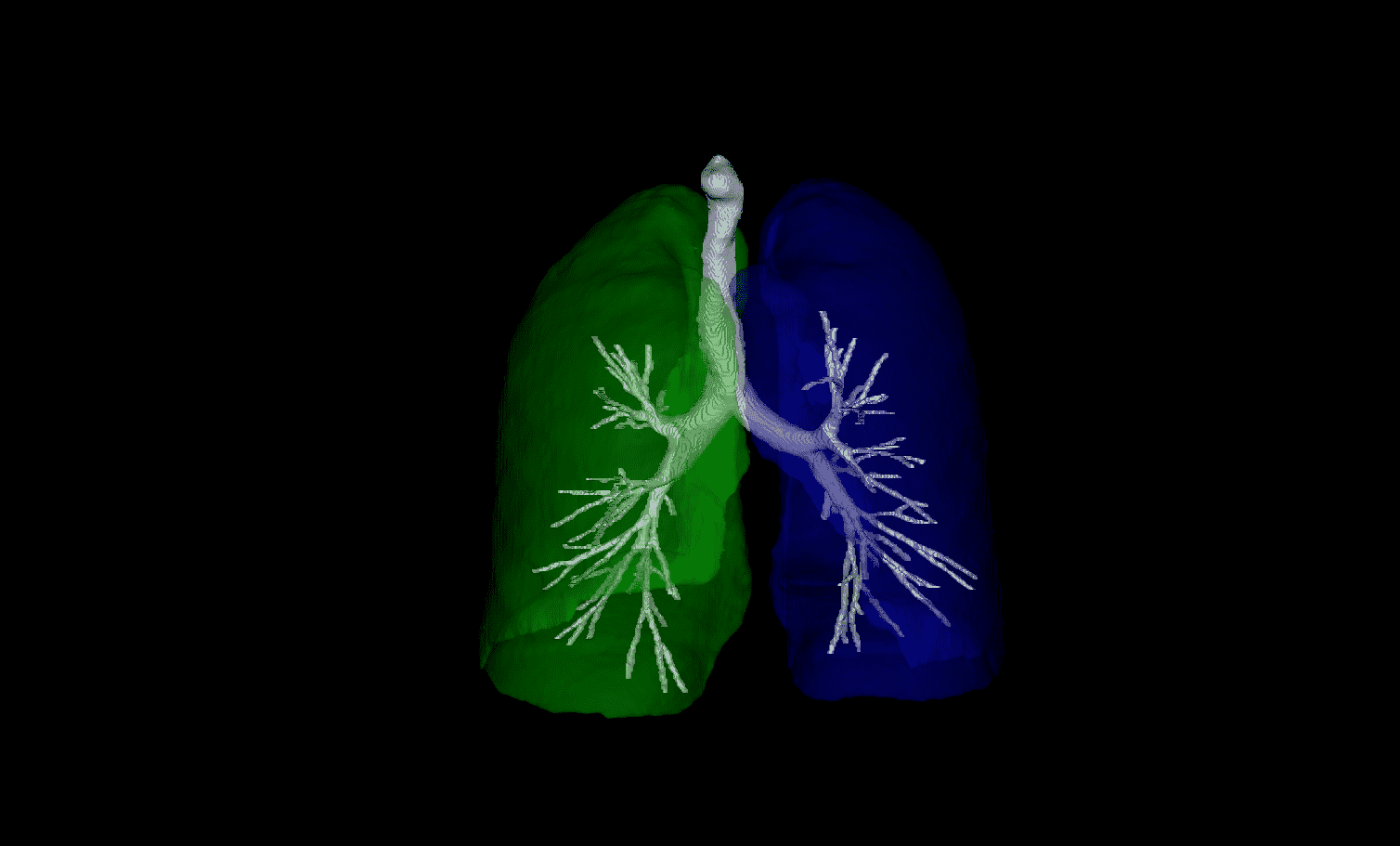}
\centerline{\small (f) Surface rendering of ours }
\end{minipage}%

\caption{Visual comparisons on the testing data for lung segmentation.
(a)-(e) show the results of the state-of-the-art methods and the proposed method, respectively.
(f) is the 3D surface rendering of lung segmented by our method.
}
\label{fig_test_lung}
\end{figure*}

\begin{figure*}[t]
\centering

\begin{minipage}[h]{\Lwid\linewidth}
\centering
\includegraphics[trim=200 10 100 60, clip, height=\Imagewidth]{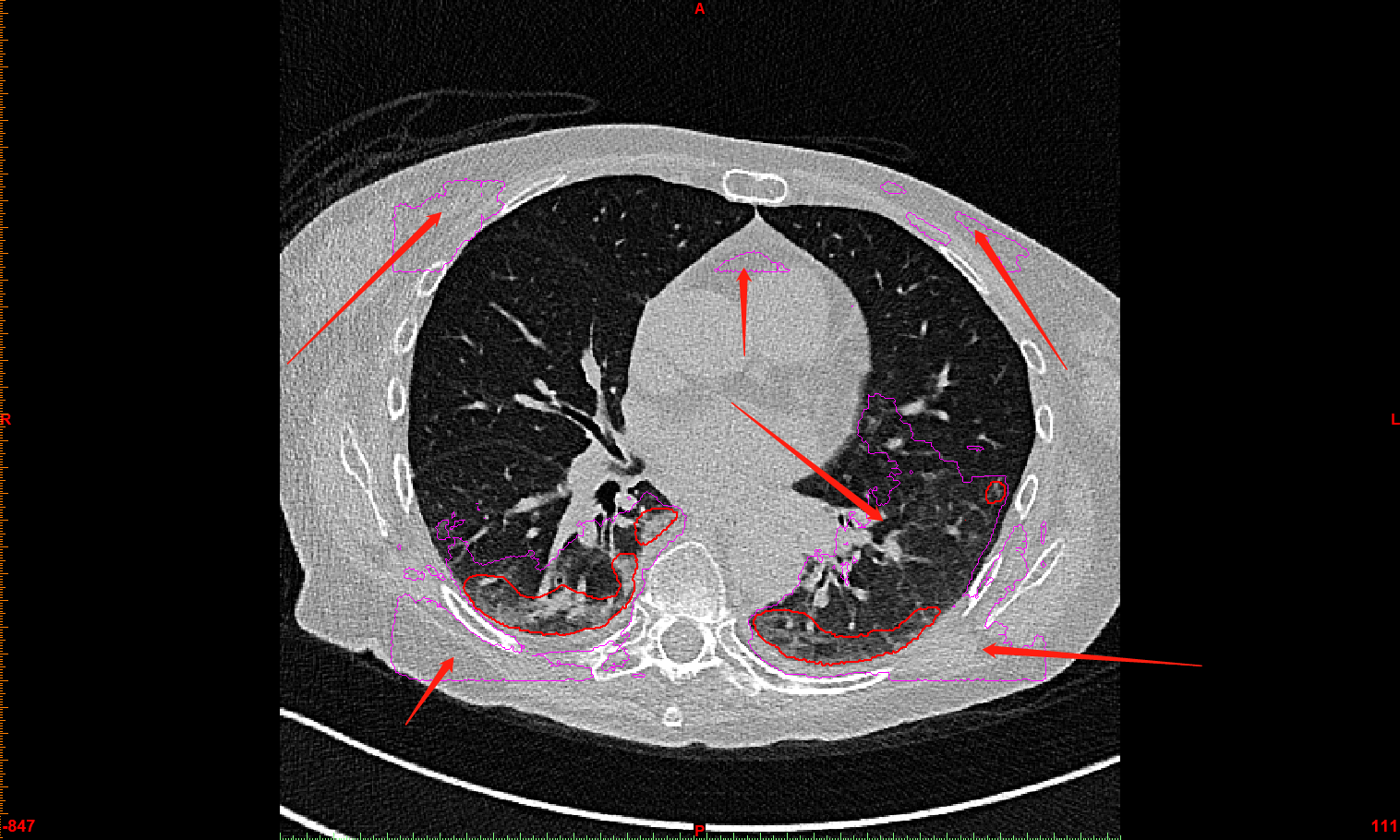}
\centerline{\small (a) FCN \cite{yang2019fd}}
\end{minipage}%
\begin{minipage}[h]{\Lwid\linewidth}
\centering
\includegraphics[trim=200 10 100 60, clip, height=\Imagewidth]{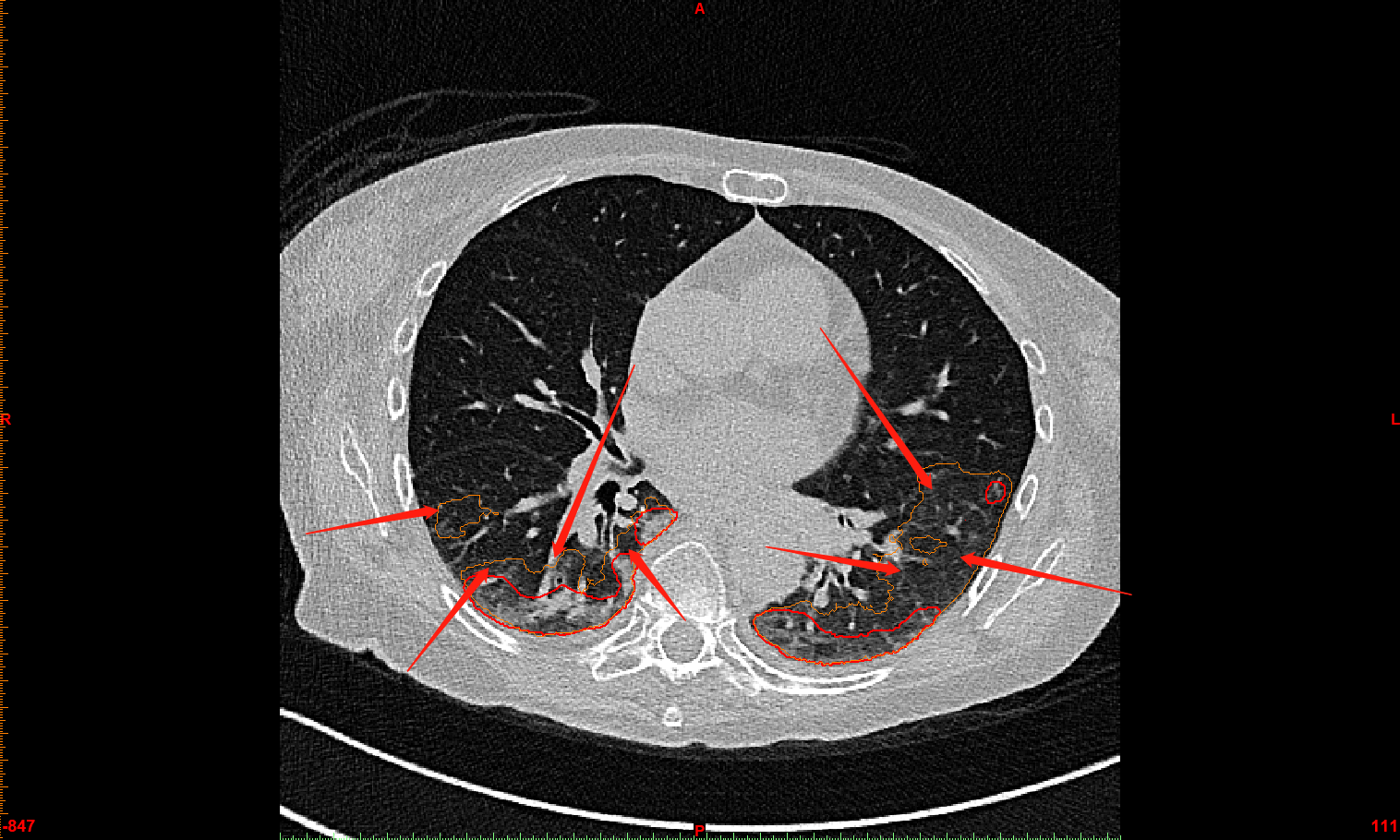}
\centerline{\small (b) UNet \cite{cciccek20163d}}
\end{minipage}%
\begin{minipage}[h]{\Lwid\linewidth}
\centering
\includegraphics[trim=200 10 100 60, clip, height=\Imagewidth]{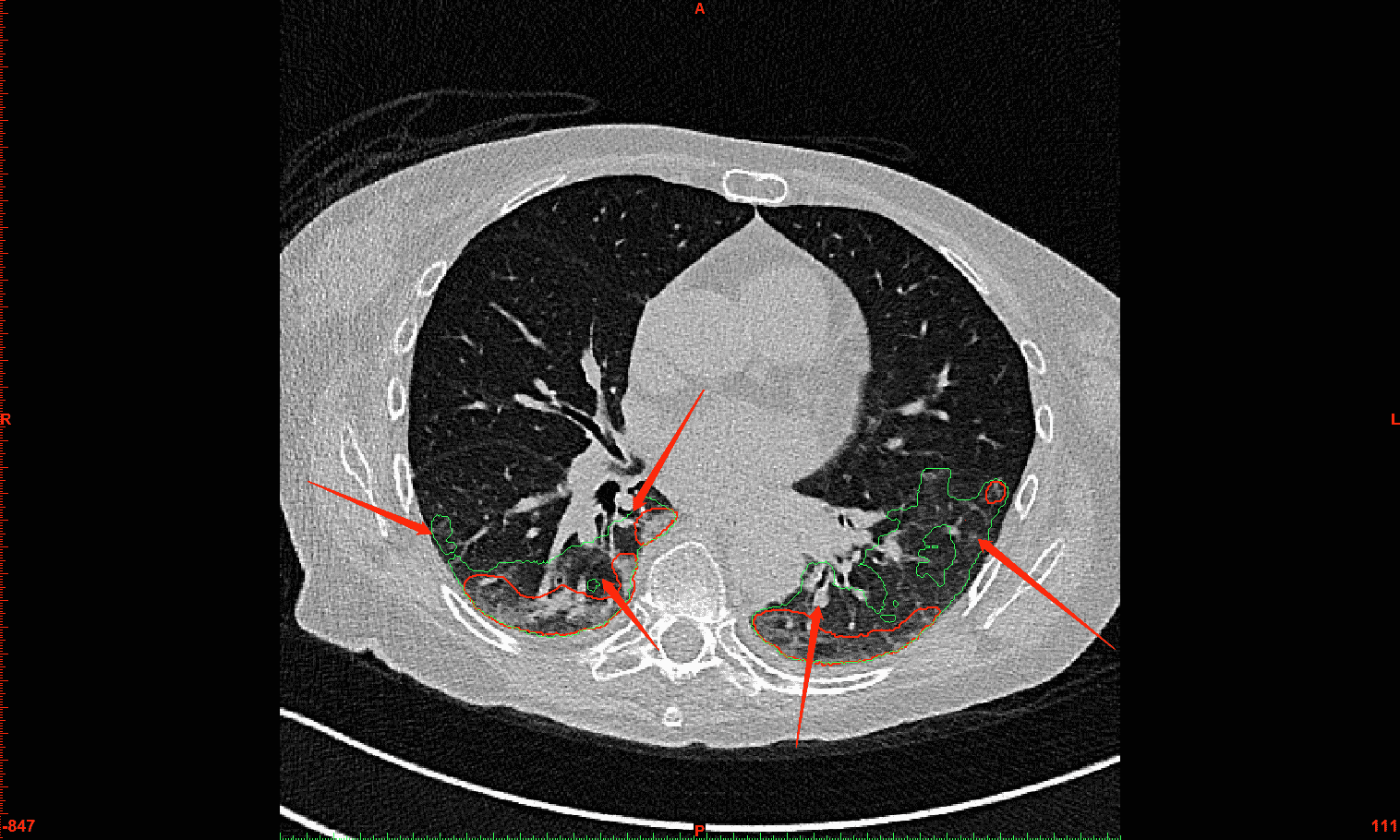}
\centerline{\small (c) UNet++ \cite{zhou18deep}
}
\end{minipage}%

\begin{minipage}[h]{\Lwid\linewidth}
\centering
\includegraphics[trim=200 10 100 60, clip, height=\Imagewidth]{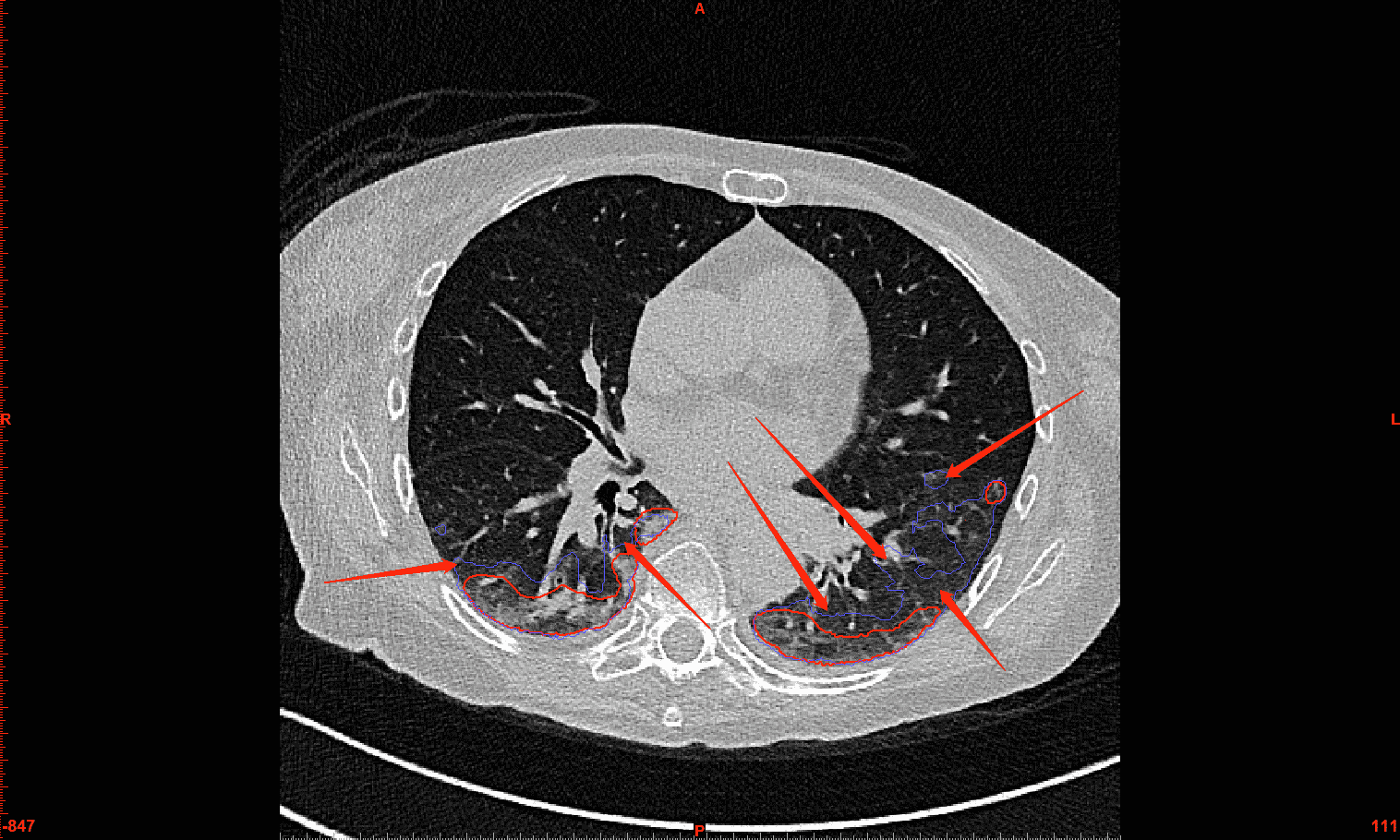}
\centerline{\small (d) VNet \cite{milletari2016v} }
\end{minipage}%
\begin{minipage}[h]{\Lwid\linewidth}
\centering
\includegraphics[trim=200 10 100 60, clip, height=\Imagewidth]{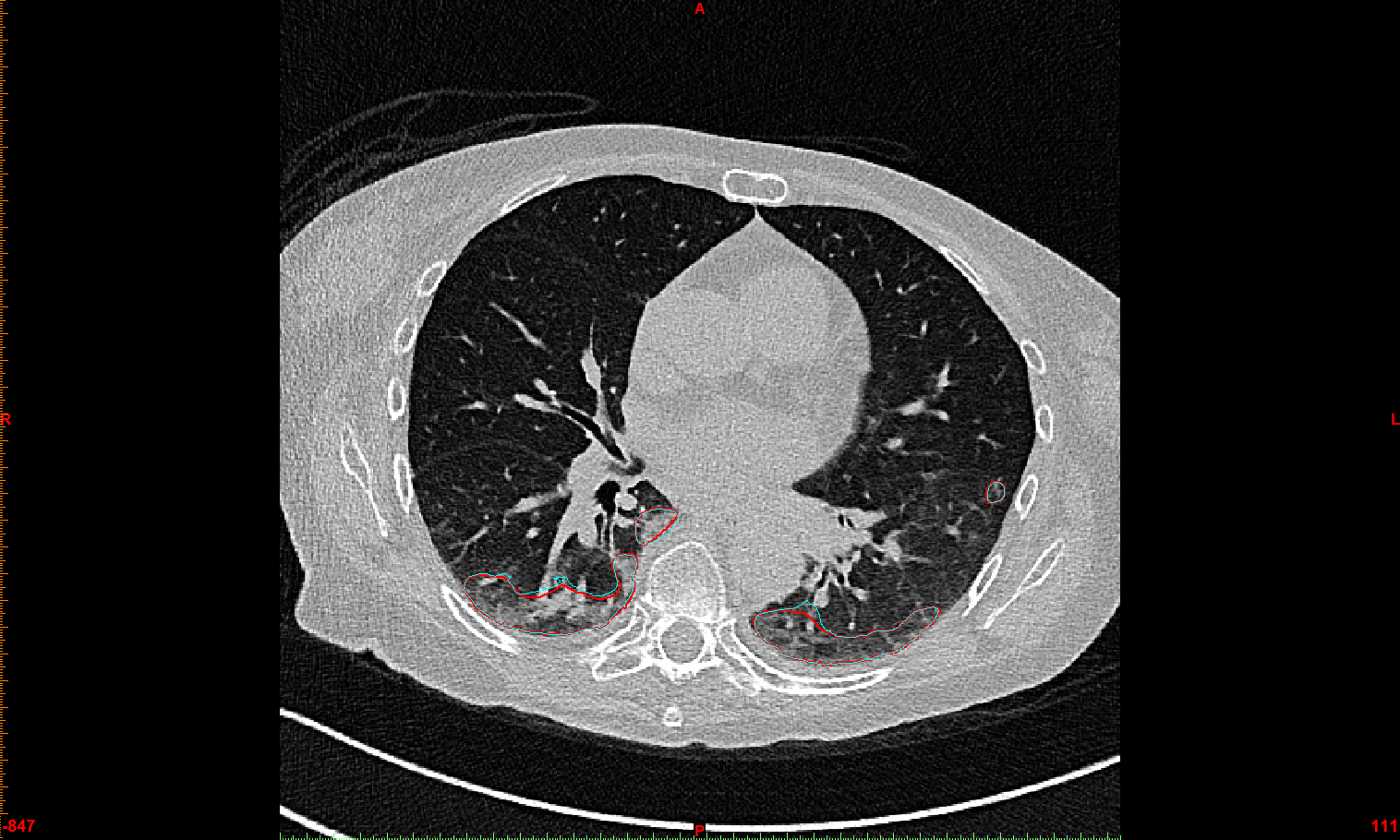}
\centerline{\small (e) Ours }
\end{minipage}%
\begin{minipage}[h]{\Lwid\linewidth}
\centering
\includegraphics[trim=200 10 100 60, clip, height=\Imagewidth]{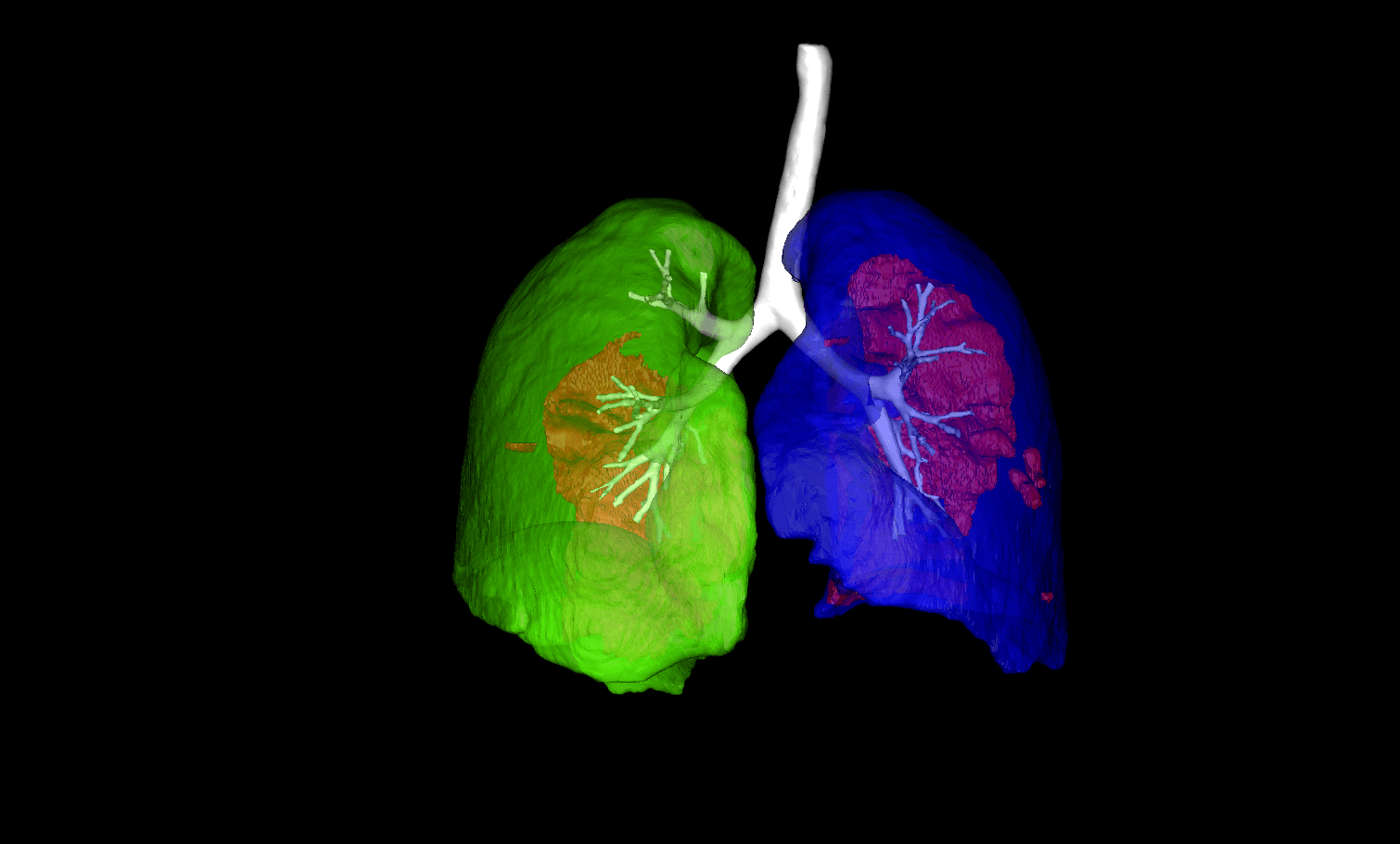}
\centerline{\small (f) Surface rendering of ours }
\end{minipage}%

\caption{Comparisons on the chest CT example of non-severe infection COVID-19 on the Germany data. The red arrows indicate the flows of different methods. Ground truth is shown with the red line, other methods are displayed with different colors.}
\label{fig_ger_les}
\end{figure*}

\begin{figure*}[t]
\centering

\begin{minipage}[h]{\Lwid\linewidth}
\centering
\includegraphics[trim=200 10 100 60, clip, height=\Imagewidth]{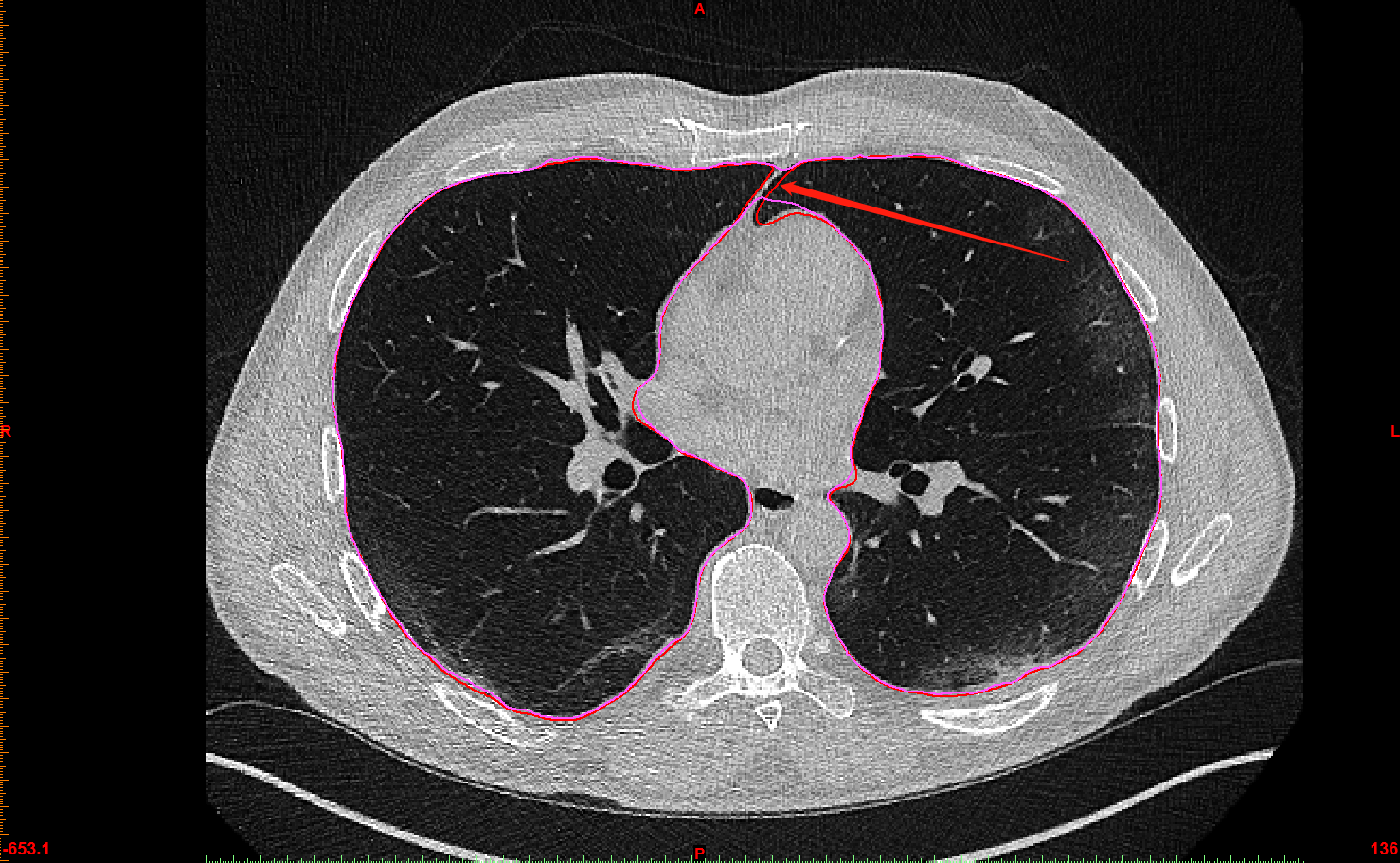}
\centerline{\small (a) FCN \cite{yang2019fd}}
\end{minipage}%
\begin{minipage}[h]{\Lwid\linewidth}
\centering
\includegraphics[trim=200 10 100 60, clip, height=\Imagewidth]{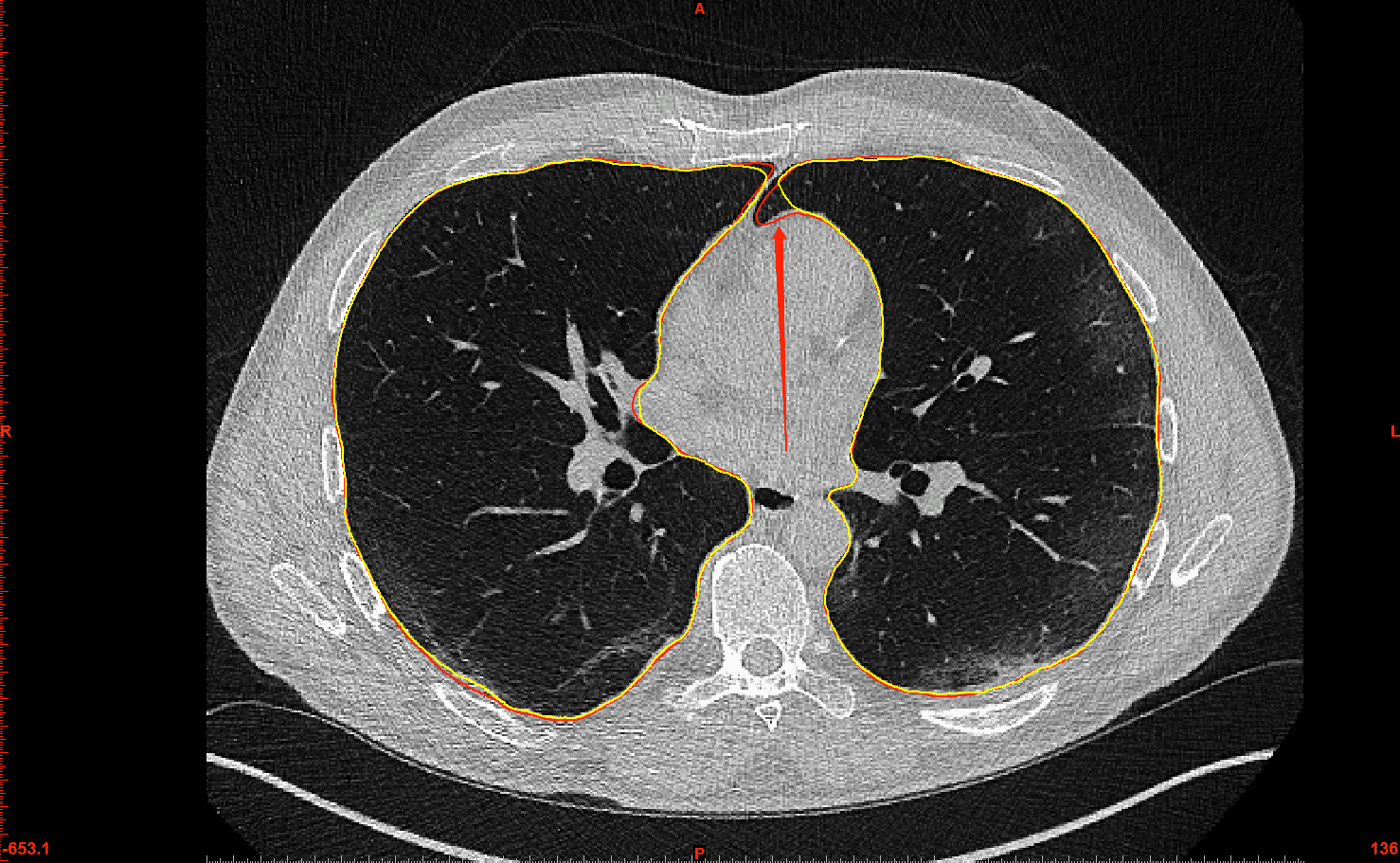}
\centerline{\small (b) UNet \cite{cciccek20163d}}
\end{minipage}%
\begin{minipage}[h]{\Lwid\linewidth}
\centering
\includegraphics[trim=200 10 100 60, clip, height=\Imagewidth]{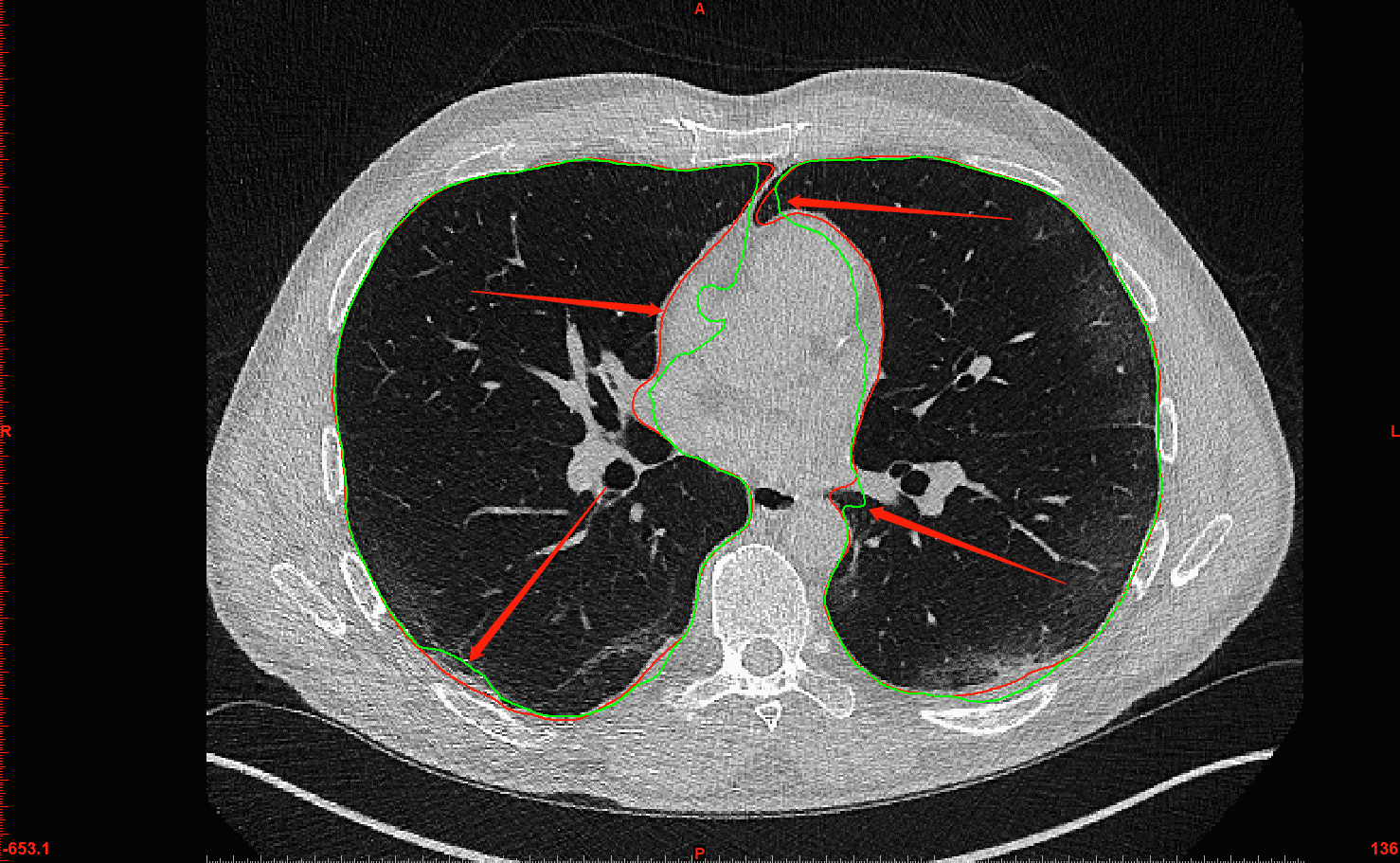}
\centerline{\small (c) UNet++ \cite{zhou18deep}
}
\end{minipage}%

\begin{minipage}[h]{\Lwid\linewidth}
\centering
\includegraphics[trim=200 10 100 60, clip, height=\Imagewidth]{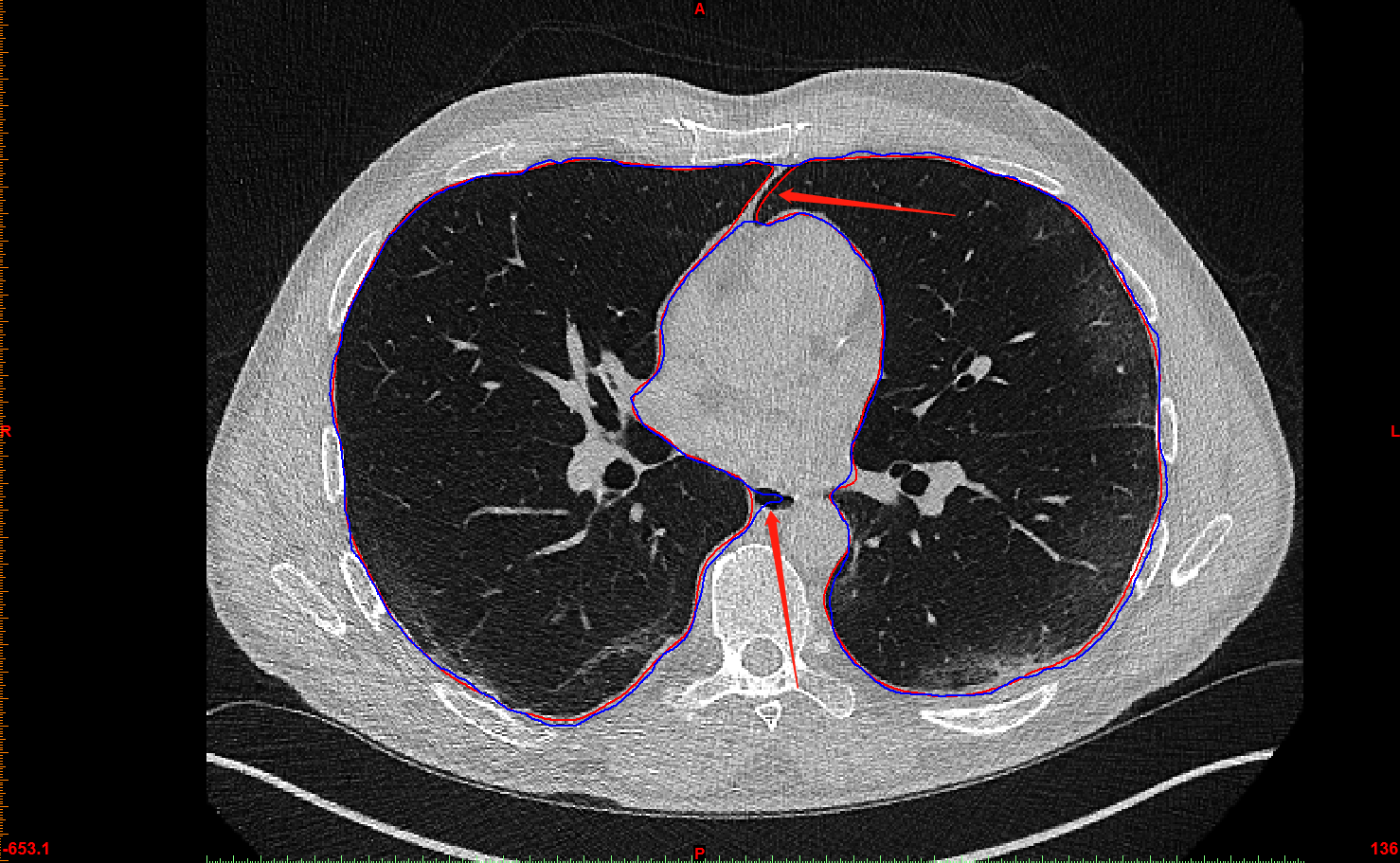}
\centerline{\small (d) VNet \cite{milletari2016v} }
\end{minipage}%
\begin{minipage}[h]{\Lwid\linewidth}
\centering
\includegraphics[trim=200 10 100 60, clip, height=\Imagewidth]{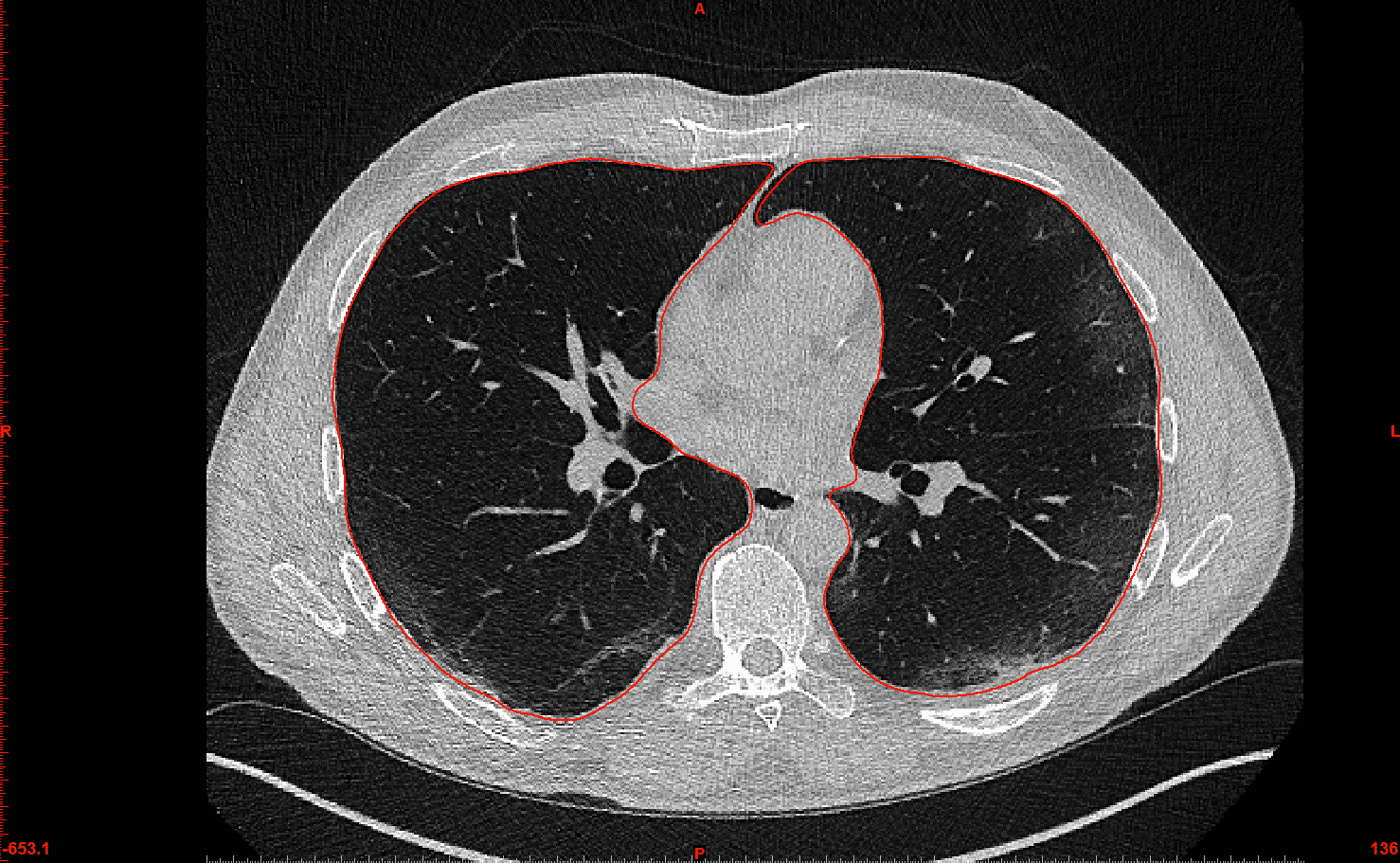}
\centerline{\small (e) Ours }
\end{minipage}%
\begin{minipage}[h]{\Lwid\linewidth}
\centering
\includegraphics[trim=200 10 100 60, clip, height=\Imagewidth]{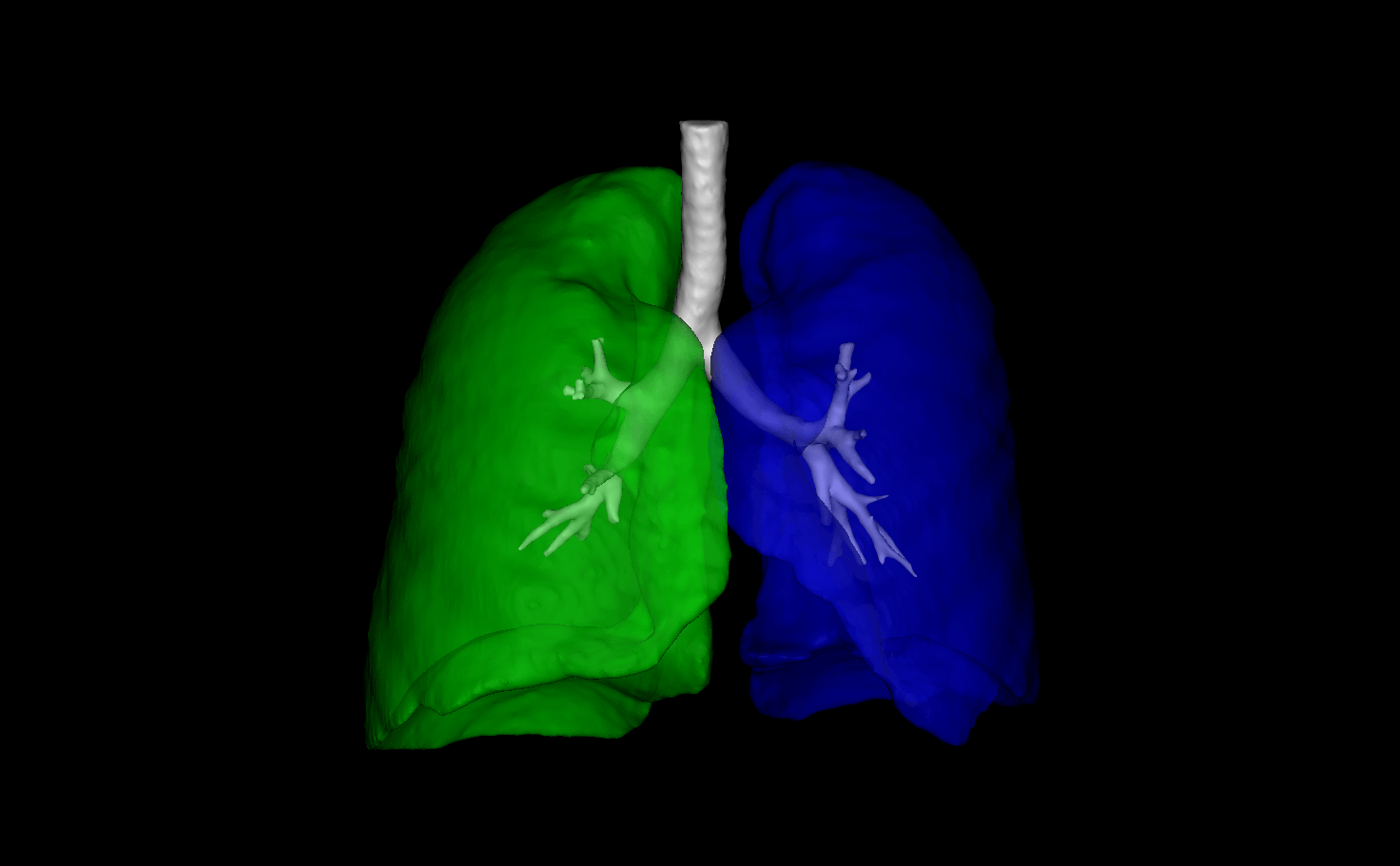}
\centerline{\small (f) Surface rendering of ours }
\end{minipage}%

\caption{The lung segmentation results of different methods on the Germany data. The red arrows indicate the flows of different methods. Ground truth is shown with the red line, other methods are displayed with different colors.}
\label{fig_ger_lun}
\end{figure*}

\subsection{Comparison with the State-of-the-art Methods}
We compare our COVID-SegNet against the previous state-of-the-art methods on two datasets (the collected domestic test set and Germany data).
Specifically, we evaluate the proposed method with FCN \cite{yang2019fd}, UNet \cite{cciccek20163d}, VNet \cite{milletari2016v} and UNet++ \cite{zhou18deep}.
Note that all methods employ 3D convolution in the framework.
The same training dataset and setting are used for all methods.

\subsubsection{Qualitative Results on the Domestic Datasets}
We compare our method with several state-of-the-art methods on the test set (Fig \ref{fig_test_1}, \ref{fig_test_2} and \ref{fig_test_3}), which contains some challenging  samples with different contrast and pathogenic conditions.

$\bullet$ \textit{COVID-19 segmentation task:}
Fig. \ref{fig_test_1} (a)-(e) illustrate the results of different methods, red line denotes the COVID-19 segmentation result of ground truth.
Since the contrast (COVID-19 and lung) of this case is not enough, these methods cannot obtain approving results.
The FCN method cannot obtain the whole edge of COVID-19.
The results of UNet++ and VNet are often scattered and overlook the overall structures of COVID-19.
The proposed method and UNet achieve better results; however, UNet result products flaw in the center of the lung (white points in (b)).
Since the proposed method employs FV blocks which adaptively enhance the global contrast of features, the proposed method can avoid the scattered artifacts.
In addition, the PASPP blocks further improve the performance of our method.
Fig \ref{fig_test_1} (f) represents the 3D surface rendering of COVID-19 infection regions segmented by our method.

Fig. \ref{fig_test_2} (a)-(e) display the example of low contrast CT images, COVID-19 infection regions are similar with chest wall.
Most of the methods can obtain massive structures of COVID-19.
However, the proposed method generates a more reasonable edge for infection regions due to the contributions of FV blocks.
Fig. \ref{fig_test_3} shows a different case captured from a non-severe patient, but the COVID-19 infection regions still hard to distinguish from the chest wall.
Thus, the methods of FCN, UNet++, VNet generate dissatisfying results.
The proposed method combined global and local information effectively obtains well-pleasing segmentation results for COVID-19 infection.

$\bullet$ \textit{Lung segmentation task:}
For the lung segmentation task, we test the performance of the proposed network on the test set.
As shown in Fig. \ref{fig_test_lung}, (a)-(b) display the results of different methods, (f) is the 3D surface rendering of our method.
From Fig. \ref{fig_test_lung}, we can easily observe that all results can close to the precision like manually annotated.
UNet++ method often miss the boundary of the lung. VNet method cannot generate a smooth margin for the lung segmentation.

\subsubsection{Qualitative Results on the Germany Data}
To verify the generalization ability of all methods, we use ten cases of data captured from Brainlab Co. Ltd. in Germany to test the segmentation of COVID-19 infection and the lung.

$\bullet$ \textit{COVID-19 segmentation task:}
Fig. \ref{fig_ger_les} shows the comparisons on the chest CT images on the Germany data.
The intensity of COVID-19 infection regions is very similar to that of the lung, which is a very challenging example.
As displayed in Fig \ref{fig_ger_les}, all state-of-the-art methods (\ie FCN, UNet, UNet++, VNet) generate perishing and over-segmentation.
Different from others, the proposed methods can obtain perfect results, which like a manual annotation (See Fig \ref{fig_ger_les} (e)).
The 3D surface rendering of the proposed method is shown in Fig \ref{fig_ger_les} (f), from which we can see that the small COVID-19 infection regions also can be segmented.

$\bullet$ \textit{Lung segmentation task:}
The segmentation results of all methods on the Germany data are shown in Fig \ref{fig_ger_lun}.
Most of all methods can generate a distinct outline of the lung.
However, from the regions marked with the red arrow, our method has a stronger segmentation ability than other state-of-the-art methods.
These perfect results demonstrate the effectiveness of the FV and PASPP blocks.

\begin{table}[h]
\begin{center}
\caption{Quantitative comparison between our method and others on the proposed test dataset. All values are the average across all test data.
} 
\label{tab:quan}
\begin{tabular}{c|c|c|c|c|c|c}
 \toprule[1.2pt]
 \textbf{Tasks}& \textbf{Metrics}&\textbf{FCN} & \textbf{UNet} & \textbf{VNet} & \textbf{UNet++} & \textbf{Ours}\\
  \hline
   \multirow{3}{*}{\textbf{COVID-19}} &  Dice & 0.659 & 0.688& 0.625& 0.681 & \textbf{0.726}\\
&Sensitive & 0.719 & 0.736 & 0.744 & 0.735 & \textbf{0.751}\\
&Precision & 0.597 & 0.662 & 0.603 & 0.719 & \textbf{0.726}\\
 \hline
 \multirow{3}{*}{\textbf{Lung}} &  Dice & 0.865 & \textbf{0.987} & 0.983 & 0.986 & \textbf{0.987}\\
&Sensitive & 0.986 & 0.987 & 0.974 & \textbf{0.988} & 0.986  \\
&Precision & 0.983 & 0.984 & 0.989 & 0.985 & \textbf{0.990}\\

 \bottomrule[1.2pt]
\end{tabular}
\end{center}
\end{table}

\begin{table*}[t]
\begin{center}
\caption{Performance of the network with different blocks.
} 
\label{tab:study}
\begin{tabular}{c|cccc|ccc|ccc|ccc}
 \toprule[1.2pt]
 \multicolumn{8}{>{\columncolor{rgb3}}c}{\textbf{Blocks}} & \multicolumn{3}{>{\columncolor{rgb6}}c}{\textbf{COVID-19 Lesion}} & \multicolumn{3}{>{\columncolor{rgb9}}c}{\textbf{Lung Segmentation}}  \\
 \hline
 \textbf{UNet4} & \textbf{CAB} & \textbf{CEB} & \textbf{PSB} & \textbf{FV} & \textbf{ASPP} & \textbf{ResASPP} & \textbf{PASPP} & \textbf{Dice} & \textbf{Sensitive} & \textbf{Precision} & \textbf{Dice} & \textbf{Sensitive} & \textbf{Precision}\\
 \hline
 $\surd$&&&&&&&&0.658& 0.670& 0.651& 0.959 & 0.956& 0.951 \\
 \hline
 $\surd$& $\surd$  &&&&&&&  0.675 & 0.683 & 0.665& 0.960 & 0.954 & 0.956 \\
 $\surd$&& $\surd$ &&&&&& 0.682  & 0.692 & 0.674 & 0.966 & 0.961  & 0.970 \\
 $\surd$&&&$\surd$ &&&&& 0.684  & 0.695 & 0.677 & 0.962 & 0.963   & 0.969 \\
 $\surd$&&&&$\surd$&&& & 0.708  & 0.729  & 0.704 & 0.975 &  0.970   & 0.981\\
 \hline
 $\surd$&&&&& $\surd$&&& 0.663  & 0.684& 0.672 & 0.968 & 0.965  &  0.971\\
 $\surd$&&&&&& $\surd$&& 0.679  & 0.701& 0.681 & 0.977 & 0.973  &  0.975\\
 $\surd$&&&&&&& $\surd$& 0.711  & 0.732& 0.707& 0.980 & 0.982  &  0.983 \\
 \hline
 $\surd$&& & & $\surd$&&&$\surd$& \textbf{0.726} & \textbf{0.751} & \textbf{0.726} & \textbf{0.987} & \textbf{0.986} & \textbf{0.990} \\
 \bottomrule[1.2pt]
\end{tabular}
\end{center}
\end{table*}

\subsubsection{Quantitative Results}
Based on the ground truths manually contoured by the radiology experts, we conduct the evaluations and comparisons to evaluate the accuracy of segmentation quantitatively.
The results are reported in Table \ref{tab:quan}, which includes lung segmentation and COVID-19 infection segmentation.

For the segmentation of COVID-19, as shown in Table \ref{tab:quan}, the results of the proposed method achieves best in all the metrics.
Thanks to the FV and PASPP block, the COVID-SegNet can effectively segment COVID-19 infection regions and significantly improve the segmentation performance over the UNet by 3.8\% in term of Dice.
All these metrics demonstrate the effectiveness of our model.

For the lung segmentation task, the average Dice similarity coefficient is 0.987.
The average sensitivity and precision are 0.986 and 0.990, respectively.
Although the existing methods have achieved enough promotion and the performance is hard to improve, the proposed COVID-SegNet still surpasses state-of-the-art methods on the term of precision.
We consider these results are attributed to the contributions of the proposed FV and PASPP blocks.

\subsection{Ablation Studies}

As shown in Table \ref{tab:study}, the baseline model is a UNet structure with 4 layers in the encoder.
We conduct the contrast enhancement branch (CEB), position-sensitive branch (PSB), and FV block, respectively.
In addition, we also replace CEB with the original channel attention block (CAB, removed the global parameter in CEB) to verify the function of global contrast enhancement.
For verifying the PASPP block, we use ASPP and ResASPP, which removes the concatenation in PASPP to prove the advantage of possessively fusing features.

\subsubsection{Study on the FV block}
The quality of the FV block, which is the combination of the contrast, global and position information, is critical for enhancing the ability of accurate COVID-19 segmentation. In this section, we first evaluate the performance of the contrast enhancement branch (CEB) from both lung and COVID-19 segmentation. Then, we study the function of the position sensitive branch (PSB).
All the comparisons are both preformed on two tasks (lung and COVID-19 segmentation).
All the results in Table \ref{tab:study} demonstrate the effectiveness of the FV blocks.

Context information is of great significance for segmenting the confusing boundary and position of COVID-19 infection regions.
To verify the performance of CEB, we employ the original channel attention block (CAB) to replace the CEB and PSB in the FV block.
From Table \ref{tab:study}, we can see that the ASPP improves the segmentation performance over the UNet4. The reason is that the features have redundant information.
However, the performance is further improved when we replace the CAB with CEB. Since the CAB merely learns the weights for each channel, the CEB uses global information to guide feature enhancement, which proves the ability of the CEB.

For PSB, it is actually a spatial attention module which has proved the effectiveness in many tasks. 
This branch focuses on the positions of features which are helpful to detect and segment COVID-19 infection regions.
As we expected, the network with PSB generates satisfying numerical results.
Combining these two branches in parallel, we obtain the FV block which consists of global (ECB) and local (PSB) information to improve the segmentation task.

\subsubsection{Study on the PASPP block}
PASPP consists of multiple atrous convolutional layers with different dilation rates and progressive concatenations.
In this part, we conduct experiments to study how different settings of PASPP influence the performance quantitatively.
We compare the PASPP block with original ASPP and modified ResASPP (removed progressive concatenations).
The results are reported in Table \ref{tab:study}, from which we obtain several conclusions.
\textit{First}, progressively fusing strategy is very effective for COVID-19 segmentation.
We deem the reason is different scale features should not be fused at once for the sophisticated COVID-19 segmentation.
With the progressively fusing, the adjacent information can better supplement the missing details.
\textit{Second}, compared with ASPP and ResASPP, sine the ResASPP includes residual learning, it obtains reasonably high performances.
This implies that the information from the early blocks can quickly flow to the output of atrous convolutional layers, and the gradient can be quickly back-propagated to the early blocks from the atrous convolutional layers.
\textit{Third}, the ASPP significantly improves the segmentation performance over the UNet4.

In general, to extract compacted features and obtain semantic information from COVID-19 CT images, we insert FV blocks into the encoder and employ PASPP for enlarging the receptive fields.
As reported in Table \ref{tab:study}, the proposed network not only achieves the best performance on lung segmentation but also on COVID-19 segmentation.

\section{Conclusion}
\label{conclusion}
In this paper, we designed and evaluated a three-dimensional deep learning model, called COVID-SegNet, for segmenting lung and COVID-19 from chest CT images.
Inspired by contrast enhancement methods and ASPP, the proposed network includes feature variation and progressive ASPP blocks, which are beneficial to highlight the boundary and position of COVID-19 infections.
These results demonstrate that the convolutional network based deep learning technology has the ability to segment COVID-19 from CT images.
We were able to collect a large number of CT images from 5 hospitals, which included 861 patients with confirmed COVID-19.
More importantly, we manually annotated these data by senior annotators.
These contributions prove the prospect of improving diagnosis and treatment for COVID-19.
In the future, we will extend the number of CT images form patients through multi-center collaborations.

\bibliographystyle{IEEEtran}
\bibliography{ref}

\end{document}